%% file: main.tex
\documentclass[12pt]{article}

\usepackage[centertags]{amsmath}
\usepackage{amsfonts}
\usepackage{amssymb}
\usepackage{mathtools} 
\usepackage{authblk}
\usepackage{bm} 
\usepackage{soul}

\usepackage{graphicx}

\usepackage{longtable} 

\usepackage{booktabs}
\usepackage{enumerate}
\usepackage{multirow}
\usepackage{floatrow} 
\usepackage[export]{adjustbox}

\usepackage{natbib} 

\usepackage{setspace}

\usepackage[tableposition=top]{caption}
\usepackage{algorithm}
\usepackage[noend]{algpseudocode}

\input{myCommands.sty}
\newtheorem{proposition}{Proposition}
\newtheorem{lemma}{Lemma}
\newtheorem{definition}{Definition}
\newtheorem{conditions}{Conditions}

\usepackage{xr-hyper} 
\usepackage{hyperref}

\def\spacingset#1{\renewcommand{\baselinestretch}%
{#1}\small\normalsize} \spacingset{1}

\makeatletter
\def\moverlay{\mathpalette\mov@rlay}
\def\mov@rlay#1#2{\leavevmode\vtop{%
   \baselineskip\z@skip \lineskiplimit-\maxdimen
   \ialign{\hfil$\m@th#1##$\hfil\cr#2\crcr}}}
\newcommand{\charfusion}[3][\mathord]{
    #1{\ifx#1\mathop\vphantom{#2}\fi
        \mathpalette\mov@rlay{#2\cr#3}
      }
    \ifx#1\mathop\expandafter\displaylimits\fi}
\makeatother

\newcommand{\blind}{0}

\begin{document}

\if0\blind
{
  \title{\bf
  Quantile Slice Sampling}
  \author[1]{Matthew~J.~Heiner}
  \author[1]{Samuel~B.~Johnson}
  \author[1]{Joshua~R.~Christensen}
  \author[1]{David~B.~Dahl}
  \affil[1]{Department of Statistics\\ Brigham Young University, Provo, Utah}
  \maketitle
} \fi

\if1\blind
{
  \bigskip
  \bigskip
  \bigskip
  \begin{center}
    {\LARGE\bf }
\end{center}
  \medskip
} \fi

\bigskip

\begin{abstract}
We propose and demonstrate a novel, effective approach to slice sampling. 
Using the probability integral transform, we first generalize Neal's shrinkage algorithm, standardizing the procedure to an automatic and universal starting point: the unit interval. 
This enables the introduction of approximate (pseudo-) targets through the factorization used in importance sampling, a technique that popularized elliptical slice sampling,  while still sampling from the correct target distribution.  
Accurate pseudo-targets can boost sampler efficiency by requiring fewer rejections and by reducing skewness in the transformed target. 
This strategy is effective when a natural, possibly crude approximation to the target exists. 
Alternatively, obtaining a marginal pseudo-target from initial samples provides an intuitive and automatic tuning procedure. 
We consider two metrics for evaluating the quality of approximation; each can be used as a criterion to find an optimal pseudo-target or as an interpretable diagnostic. 
We examine performance of the proposed sampler relative to other popular, easily implemented MCMC samplers on standard targets in isolation, and as steps within a Gibbs sampler in a Bayesian modeling context. 
We extend the transformation method to multivariate slice samplers and demonstrate with a constrained state-space model for which a readily available forward-backward algorithm provides the target approximation. 
Supplemental materials 
 and accompanying R package \texttt{qslice} 
are available online.
\end{abstract}

\noindent%

\newpage

\spacingset{1.7} 


\input{sec_introduction.tex}
\input{sec_transformation.tex}
\input{sec_approx_targets.tex}
\input{sec_selecting_pseudo_target.tex}
\input{sec_illustrations.tex}
\input{sec_multivariate.tex}
\input{sec_DHR.tex}
\input{sec_discussion.tex}

\section*{Acknowledgements}
The authors gratefully acknowledge helpful conversations with and suggestions from Alejandro Jara, Benjamin Dahl, Godwin Osabutey, Garritt Page, Richard Warr, and Carlos Sing Long, as well as valuable suggestions from anonymous reviewers.

%

\bigskip
\begin{center}
{\large\bf SUPPLEMENTARY MATERIALS}
\end{center}

\begin{description}

\item[Appendix:] (.pdf file)
Proofs of propositions in Section \ref{sec:transform}; 
proof of Proposition \ref{prop:sliceWidth}; 
additional details for the simulation studies; 
additional details for the $g$-prior illustration; 
proof of Proposition \ref{prop:multivariate_unif_ergodicity};
implementation details for multivariate illustration. 

\item[Code:] (GitHub repository) \url{https://github.com/mheiner/qslice_examples}. \\ 
R scripts to recreate examples and results in the paper (see README files). 

\end{description}

\bibliography{references.bib}

\newpage

\newcommand{\beginsupplement}{
        \setcounter{section}{0}
        \renewcommand{\thesection}{A\arabic{section}}
        \setcounter{table}{0}
        \renewcommand{\thetable}{A\arabic{table}}
        \setcounter{figure}{0}
        \renewcommand{\thefigure}{A\arabic{figure}}
        \setcounter{page}{1}
        \renewcommand{\thepage}{A\arabic{page}}
        \setcounter{equation}{0}
        \renewcommand{\theequation}{A\arabic{equation}}
        \setcounter{algorithm}{0}
        \renewcommand{\thealgorithm}{A\arabic{algorithm}}
        \setcounter{definition}{0}
        \renewcommand{\thedefinition}{A\arabic{definition}}
        \setcounter{lemma}{0}
        \renewcommand{\thelemma}{A\arabic{lemma}}
        \setcounter{proposition}{0}
        \renewcommand{\theproposition}{A\arabic{proposition}}
        \setcounter{conditions}{0}
        \renewcommand{\theconditions}{A\arabic{conditions}}

}

\beginsupplement

\input{appx_proofs_transform.tex}
\input{appx_proof_sliceWidth.tex}
\input{appx_simulation_details.tex}
\input{appx_gprior.tex}
\input{appx_proofs_multivariate.tex}
\input{appx_DHR.tex}

\end{document}

%% file: sec_introduction.tex
\section{Introduction}
\label{sec:intro}

The landmark paper of \citet{neal2003slice} brought slice sampling into the mainstream of Markov chain Monte Carlo (MCMC) methods by introducing an automatic procedure for the difficult step of sampling within an unknown ``slice region." 
In this article, we extend the core component of that procedure to work with more general slice samplers, enabling its use with a reformulation of the slice sampler 
that can substantially improve performance. 
The reformulation introduces a user-defined proxy $\hat\pi$ to the target density $\pi(\theta) \propto g(\theta)$ via the factorization used in importance sampling, 
\begin{align}
    \label{eq:importance}
    \begin{split}
    g(\theta) &= \frac{g(\theta)}{\hat{\pi}(\theta)} \hat{\pi}(\theta) \\
    &= h(\theta) \hat{\pi}(\theta) \, ,
    \end{split}
\end{align}
an approach adopted by \citet{nishihara2014parallellipslice} to broaden the scope of elliptical slice sampling \citep{murray2010elliptical} beyond targets containing Gaussian kernels. 
Aided by the probability integral transform, our extended procedure and its application in \eqref{eq:importance} together comprise what we call \textit{quantile slice sampling}.

Slice sampling \citep{swendsen1987slice, edwards1988slice, besag1993, neal2003slice,higdon1998auxiliary, damlen1999gibbs} decomposes, into manageable steps, the task of drawing from a distribution that is difficult to sample directly. 
Consider real-valued $\theta$ with target density $\pi(\theta) \propto g(\theta) = \mathcal{L}(\theta) f(\theta)$, factored such that we can easily sample from the distribution having density $f(\theta)$. 
Augmenting with a latent auxiliary variable $V$ and specifying the joint density of $\theta$ and $V$ to be proportional to $f(\theta) \, 1\{0 < v < \mathcal{L}(\theta)\}$ (where $1\{\cdot\}$ returns 1 for true statements and 0 otherwise), we recover a marginal density of $\theta$ proportional to $g(\theta)$. 
Given $\theta$, $V$ is uniformly distributed on the interval $(0, \mathcal{L}(\theta))$. 
Given $V = v$, $\theta$ has density $f$ with support restricted to $A_\mathcal{L}(v) \defeq \{ \theta : v < \mathcal{L}(\theta) \}$, commonly known as the slice region. 
This augmentation admits convenient Gibbs sampling in what \citet{roberts1999convergence} call the \textit{simple} slice sampler. 

Despite their excellent theoretical properties \citep{roberts1999convergence, mira2002sliceEfficiency, planas2018symmetry}, simple slice samplers are limited by the challenge of finding the slice region, which is specific to the target density. 
If $A_\mathcal{L}(v)$ can be found analytically, sampling requires a custom solution. 
If $A_\mathcal{L}(v)$ is not known, \textit{hybrid} strategies embed Markov transition kernels in the second full conditional update (of $\theta \mid V = v)$ to obtain a draw restricted to $A_\mathcal{L}(v)$ \citep{roberts1997geometric, rudolf2018hybrid, latuszynski2024hybrid, power2024hybridslice}. 
Many hybrid samplers require multiple evaluations of the target density and some require tuning. 

\citet{neal2003slice} introduced a general class of hybrid techniques to partially automate the search for slice region $A_g(v)$ in \textit{uniform} slice samplers (that target $g$ directly by alternating draws $V \mid \theta$ uniformly on $(0, \, g(\theta))$ and $\theta \mid V = v$ uniformly over $A_g(v) = \{ \theta : v < g(\theta) \}$). 
Central to these techniques is the \textit{shrinkage procedure}, a step that adaptively shrinks an initial interval thought to contain $A_g(v)$ by rejecting proposed draws that lie outside the slice region. 
It was designed only for uniform full-conditional distributions, and it is typically supported by preliminary steps to set the initial interval, which is redefined at each iteration of MCMC. 

Our reformulation of the simple slice sampler provides a natural environment for the shrinkage procedure. 
Use of the probability integral transform reveals an automatic, generalized version of 
the shrinkage procedure that extends its application from uniform to simple slice sampling. 
This enables introduction of a single user input: 
an approximate target density $\hat{\pi}(\theta)$ 
that enters via the factorization in \eqref{eq:importance}. 
Simple slice sampling then proceeds by defining the slice with the importance ratio $h(\theta) = g(\theta) / \hat{\pi}(\theta)$ and drawing from $\hat{\pi}(\theta)$ restricted to $A_h(v)$ using the generalized shrinkage procedure. 

\begin{figure}[t!]
    \centering
    \includegraphics[width = 5.4in]{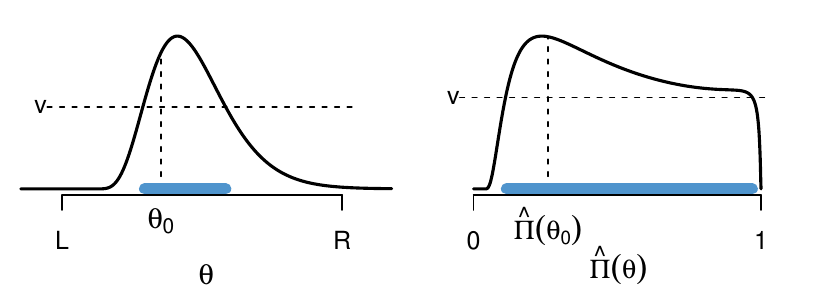}
    \caption{Illustration of the effect of transformation using a high-quality target approximation. 
    The latent $v$ that defines the slice region (thick line along axis) is at the same height relative to the target evaluation at the previous point ($\theta_0$ or $\widehat{\Pi}(\theta_0)$) in both the original target (left) and target transformed according to the approximating distribution function $\widehat{\Pi}$ (right). 
    However, the slice region fills more of the initial interval in the transformed case, resulting in fewer rejections, on average.}
    \label{fig:intro_cartoon}
\end{figure}

Our quantile slice sampler, or transformation with an approximate target, is illustrated in Figure \ref{fig:intro_cartoon}. 
When a reasonably good approximate target can be found, the method replaces the problem of uniformly slice sampling the original target density in the left panel with that of uniformly slice sampling the density in the right. 
In the original setting, finding an initial interval $(L, R)$ to cover $A_g(v)$ often requires multiple target evaluations. 
In contrast, the initial interval under transformation using the distribution function of the approximate target, $\widehat{\Pi}(\theta)$, is always $(0,1)$. 
The transformed target density (right) flattens, approaching uniform as $\hat{\pi}(\theta)$ approaches $\pi(\theta)$, effectively expanding the slice region and reducing the expected number of target evaluations taken by the shrinkage procedure. 
Samples on the transformed space can also be used for interpretable diagnostics, indicating adjustments to the pseudo-target that can improve efficiency. 

Our proposed sampler can be viewed as a slice-sampling analog of the independence Metropolis-Hastings algorithm (IMH; \citealp{liu1996imh}), in which $\hat{\pi}$ acts as the proposal distribution. 
While high-fidelity approximations to the target yield the highest computational efficiency, it is not critical to optimize the approximation. 
Rather, one can use an easy-to-specify, possibly rough approximation and rely on the mechanics of hybrid slice sampling to correct discrepancies and produce samples from the correct target distribution. 
The approximate target may be fixed throughout MCMC sampling or may change at each iteration (to target, for example, full-conditional distributions within a Gibbs sampler). 
The result is a straightforward approach to slice sampling that can yield excellent performance.

This article proceeds as follows. 
First, we generalize the shrinkage procedure in Section \ref{sec:transform}. 
The generalization facilitates introducing the approximate target in Section \ref{sec:QSS}. 
We examine approaches to specifying the approximate target in Section \ref{sec:pseudo_target_selection}, where the transformation enables evaluation of approximation quality. 
Section \ref{sec:illustrations} compares empirical performance of the proposed quantile slice and other samplers on standard univariate targets as well as within a Gibbs sampler in a Bayesian modeling scenario. 
Section \ref{sec:multivariate} extends transformation and approximate targets to Neal's \citeyearpar{neal2003slice} shrinkage procedure on hyperrectangles for multivariate sampling. 
Our multivariate quantile slice sampler is then applied in Section \ref{sec:DHR} to facilitate posterior sampling in a high-dimensional setting in which traditional slice sampling performs poorly. 
We conclude with a discussion in Section \ref{sec:discussion}. 
The online appendix includes proofs and additional details. 
Software implementing quantile slice samplers for user-defined targets is available in the \texttt{qslice} R package on CRAN \citep{qslice_package, Rteam}.

%% file: sec_transformation.tex
\section{Transformation extends Neal's shrinkage procedure}
\label{sec:transform}

\citet{neal2003slice} introduced the \textit{shrinkage procedure} as a hybrid step to sample on the slice region $A_g(v)$. 
The procedure, 
given in Figure 5 of \citet{neal2003slice}, 
has an important role in recent innovations such as elliptical \citep{murray2010elliptical}, factor \citep{tibbits2014factorslice}, and latent \citep{li2023latent} slice sampling. 
Here, we generalize the shrinkage procedure with a transformation that i) extends its use from uniform slice samplers to simple slice samplers and ii) proves useful for characterizing approximation quality in Section \ref{sec:pseudo_target_selection}. 

A primary challenge to slice sampling with the shrinkage procedure selection of the initial interval, which requires care. 
Starting too narrow can result in slow exploration, or even a reducible Markov chain if the target has separated support \citep{mira2003discussion}. 
\citet{neal2003slice} introduces two procedures (called ``stepping out" and ``doubling") for initializing the interval before shrinking. 
Both require tuning or additional checks.  
\citet{li2023latent} address the initialization challenge with stochastically adaptive intervals. 
If the target distribution has bounded support, initializing the interval to match the bounds avoids the search. 
However, if the target density concentrates mass away from the bounds, many shrinkage steps will require many target evaluations, thereby increasing runtime.

We bypass the problem of selecting an initial interval with a transformation that gives automatic and universal bounds. 
Consider a random variable $\theta$ with target distribution $\Pi$ and density $\pi(\theta) \propto \mathcal{L}(\theta) f(\theta)$, factored into probability density $f(\theta)$ with Lebesgue-measurable support on $\mathcal{S}_F \subseteq \mathbb{R}$, and nonnegative, lower semicontinuous function $\mathcal{L}(\theta)$ with Lebesgue-measurable support on $\mathcal{S} \subseteq \mathcal{S}_F$. 
Let $F(\theta)$ denote the distribution function associated with $f$ and assume that its inverse $F^{-1}(u)$ can be computed for all $u \in (0,1)$. 
Here, a simple slice sampler is the natural candidate. 
Rewriting the unnormalized target as $g(\theta) \, 1\{\theta \in \mathcal{S} \}$ where $g(\theta) = \mathcal{L}(\theta) f(\theta)$ returns to the uniform case, but does not necessarily bound the support. 
Alternatively, we can return to the uniform case \textit{and} bound the support in the following way. 

Consider the transformation $\phi = F(\theta) \in \img_F(\mathcal{S}) \subseteq (0,1)$. Straightforward application of the probability integral transform yields an equivalent target 
\begin{align}
    \label{eq:transformedtarget}
    {\pi_\phi}(\phi) \propto \mathcal{L}(F^{-1}(\phi)) \Unifdist(\phi; 0, 1) \, ,
\end{align}
where $\Unifdist(\theta; a, b)$ denotes the uniform density on the interval $(a,b)$. 
This transformation admits a uniform slice sampler for which the support is always a subset of the unit interval. 
In the ideal setting with no hybrid search for the slice region, the two above slice samplers yield equivalent Markov chains; 
a proof is given in Section \ref{sec:equiv_chains} of the appendix. 
With the mild conditions of the preceding paragraph, which trivially hold in most applications, 
a simple slice sampler can be implemented as a uniform slice sampler. 
Draws $\{ \tilde{\phi}^{(s)} : s = 1, \ldots, S \}$ are transformed back to $\tilde{\theta}^{(s)} \leftarrow F^{-1}(\tilde{\phi}^{(s)})$, which has stationary distribution $\Pi$. 
 
The transformation reveals that the shrinkage procedure extends beyond the uniform case to any simple slice sampler for which the factorization of target $\pi(\theta)$ includes a density with corresponding (monotonic and computable) CDF and inverse-CDF. 
Shrinking a uniform distribution over the slice region for $\phi$ equivalently shrinks \emph{on the quantiles} of $F$. 
We next introduce the extension of the shrinkage procedure, 
first studying its properties in isolation, with no reference to slice sampling. 
It is then applied in the context of simple slice sampling, admitting equivalent Markov chains between simple and uniform samplers as before, but now involving shrinkage procedures.

\subsection{Generalization of the shrinkage procedure}
\label{sec:genshrink}

The shrinkage procedure itself provides a transition kernel for any continuous target with restricted support, separate from slice sampling. 
To sample from a distribution $F$ with support on $\mathcal{S}_F \subseteq \mathbb{R}$ but restricted to a fixed set $A \subset \mathcal{S}_F$ with $F(A) > 0$ (that is, $\Pr(\theta \in A) > 0$ under $F$), we construct a Markov chain that draws candidates from $F$ restricted to an interval (initially $(\inf \mathcal{S}_F, \sup \mathcal{S}_F)$) that shrinks around the previous value until a sample within $A$ is found. 
A single scan of the procedure is outlined in Algorithm \ref{alg:generalShrink}. 

Let $K_{Q,A}$ denote the corresponding transition kernel, with unrestricted, continuous target $Q$. 
Denote the target with restricted support as $Q_A$.
An MCMC sampler initialized at $\theta^{(0)} \in A$ that iteratively applies Algorithm \ref{alg:generalShrink} with $Q = F$ and $x_0 = \theta^{(s)}$ will converge to stationary distribution $F_A$, as established in Proposition \ref{prop:generalShrink}. 
A proof is given in Section \ref{sec:proof_generalShrink} of the appendix. 

\begin{algorithm}[tb]
\caption{Generalized shrinkage procedure.}\label{alg:generalShrink}
\begin{algorithmic}
\Require \\
$Q$: continuous distribution to be sampled with unrestricted support $\mathcal{S}_Q$ \\
$A$: subset of $\mathcal{S}_Q$ to which the support of $x$ is to be restricted \\
$x_0$: current state (in $A$)

\Ensure \\ 
$x_1$: new state \\

\medskip
\hrule

\State $L \gets \inf \mathcal{S}_Q$, $R \gets \sup \mathcal{S}_Q$
\Loop
\State Draw $x_1 \sim Q(x \mid L \le x \le R)$ \Comment{Sample $Q$ restricted to $x \in (L, R)$.}
\If{$ x_1 \in A$} \textbf{break}
\Else
    \If{$x_1 < x_0$} $L \gets x_1$
    \Else~$R \gets x_1$
    \EndIf
\EndIf
\EndLoop
\end{algorithmic}
\end{algorithm}

\begin{proposition}
    \label{prop:generalShrink}
    The Markov transition kernel $K_{Q,A}$ outlined in Algorithm \ref{alg:generalShrink} using $Q$ with continuous and monotonic distribution function on support $\mathcal{S}_Q$, and $A \subset \mathcal{S}_Q$ with $Q(A) > 0$, is $Q_A$-reversible. If initialized in $A$, the resulting Markov chain is uniformly ergodic on $A$.
\end{proposition}

Algorithm \ref{alg:generalShrink} yields an independent draw directly from $F_A$ if $A$ is a single interval, as shrinkage steps outside of $A$ have no influence on subsequent accepted draws within $A$. 
On the other hand, if $A$ consists of multiple separated intervals, it is possible that shrinking steps exclude parts of $A$, inducing autodependence and slowing convergence. 
This can occur, for example, in slice sampling with multimodal targets. 

To see how a separated $A$ can affect convergence, consider the following heuristic illustration of applying Algorithm \ref{alg:generalShrink} with $Q$ being uniform over the unit interval and $A = (0, a) \cup (b, 1)$ with $0<a<b<1$. 
Next consider a binary Markov chain that tracks whether the state is in the right interval, that is, $Z^{(t)} = I\{X^{(t)} \in (b,1)\}$. 
The transition probabilities are $p(0|0) \defeq \Pr(Z^{(t)} = 0 \mid Z^{(t-1)} = 0) = b$, $p(1|0) = 1 - b$, $p(0|1) = a$, and $p(1|1) = 1 - a$. 
The second eigenvalue of the transition matrix is a monotonically increasing function of $b - a$ only. 
Thus, increasing $b - a$ decreases the spectral gap of $\{Z^{(t)}\}$, slowing the mixing of the chain. 
Generalizing to non-uniform $Q$, mixing times increase with the probability, under $Q$, of the interval separating components of $A$.

\subsection{Shrinkage procedure within slice sampling}

Algorithm \ref{alg:generalShrink} is useful for restricting the support of a distribution to some subset $A$ with $F(A) > 0$ that is not known analytically, but for which $\theta \in A$ can be tested. 
This is the problem motivating hybrid slice sampling. 
Proposition \ref{prop:shrink_in_slice} establishes that embedding Algorithm \ref{alg:generalShrink} in a simple slice sampler as a hybrid step yields, under mild conditions, a $\pi$-invariant and ergodic Markov chain. 
A proof is given in Section \ref{sec:proof_shrinkinslice} of the appendix.

\begin{proposition}
    \label{prop:shrink_in_slice}
    Assume target density $\pi(\theta) \propto \mathcal{L}(\theta) f(\theta)$ and all conditions listed in the paragraph preceding \eqref{eq:transformedtarget}. 
    The simple slice sampler that defines the slice region with $V \mid \theta \sim \Unifdist(0, \mathcal{L}(\theta))$ and implements Algorithm \ref{alg:generalShrink} as a hybrid step targeting $p(\theta \mid V=v) \propto f(\theta) I\{ \theta \in A_\mathcal{L}(v) \}$ is $\pi$-reversible. 
    Furthermore, if $\sup_\theta \mathcal{L}(\theta) < \infty$, then this hybrid simple slice sampler is uniformly ergodic. 
\end{proposition}

Note that the generalized shrinkage procedure in Algorithm \ref{alg:generalShrink} is applied directly in the second step (targeting $\theta \mid V = v$) in the simple slice sampler for $\pi(\theta) \propto \mathcal{L}(\theta) f(\theta)$. 
Although the transformation to $\phi$ is implicit here, including transformed draws aids with intuition and with quantifying pseudo-target quality in Section \ref{sec:pseudo_target_selection}. 
Finally, a simple (or uniform) slice sampler with generalized shrinkage can also embed within a larger Gibbs sampler. 

%% file: sec_approx_targets.tex
\section{Quantile slice sampler}
\label{sec:QSS}

The transformation in Section \ref{sec:transform} simplifies working with the shrinkage procedure and extends its use beyond uniform slice sampling to the broader class of simple slice samplers. 
This enables the introduction in Section \ref{sec:approxTargets} of an approximate target to improve sampler efficiency, a strategy that has proven useful for elliptical slice sampling. 
The result is a quantile slice sampler that can be viewed as a slice-sampling analog of independence Metropolis-Hastings (IMH); 
we explore the connection in Section \ref{sec:IMH}.

\subsection{Approximate targets}
\label{sec:approxTargets}

If there is substantial disagreement between the target $\pi$ and component $f$ densities, a uniform simple slice sampler 
applied to \eqref{eq:transformedtarget} will be inefficient, often requiring several shrinkage steps. 
To improve efficiency, we can employ what \citet{li2020parallel} call a pseudo-prior. 
Suppose that density $\hat{\pi}(\theta)$ provides an approximation to $\pi$ on the same (or extended) support, and that we can evaluate its inverse CDF, $\widehat{\Pi}^{-1}(u)$, for $u \in (0,1)$. 
Following \citet{nishihara2014parallellipslice}, \citet{fagan2016elliptical},  \citet{li2020parallel}, and \citet{mira2002sliceEfficiency}, we rewrite the unnormalized target density 
\begin{align}
    \label{eq:pseudoprior}
    \begin{split}
        \pi(\theta) & \propto \mathcal{L}(\theta) f(\theta)  \\
        & = \frac{ \mathcal{L}(\theta) f(\theta) } { \hat{\pi}(\theta) } \, \hat{\pi}(\theta) \\
        & = h(\theta) \, \hat{\pi}(\theta) \, , 
    \end{split}
\end{align}
where the importance ratio $h(\theta)$ takes the place of the ``likelihood" and $\hat{\pi}(\theta)$ assumes the role of ``pseudo-prior." 
We call $\hat{\pi}$ the \textit{pseudo-target} density to accurately reflect its role in \eqref{eq:pseudoprior}. 
Note that the factorization $\mathcal{L}(\theta) f(\theta)$ is not necessary; $h(\theta)$ in \eqref{eq:pseudoprior} could be equivalently expressed as $g(\theta) / \hat{\pi}(\theta)$ for any $g(\theta) \propto \pi(\theta)$.
The slice sampler using $h(\theta)$ to define the slice and drawing from restricted $\hat{\pi}(\theta)$ is uniformly ergodic if $h(\theta)$ is bounded (Theorem 6 of \citealp{mira2002sliceEfficiency}). 
Proposition \ref{prop:shrink_in_slice} extends this result to hybrid simple slice samplers using shrinkage procedures. 
Unlike the transformation in Section \ref{sec:transform}, rewriting the target as in \eqref{eq:pseudoprior} yields a new slice sampler that defines a distinct Markov chain.

\begin{algorithm}[tb]
\caption{Our proposed quantile slice sampler.}\label{alg:quantile_slice}
\begin{algorithmic}
\Require \\
$g$: unnormalized target density function with support $\mathcal{S} \subseteq \mathbb{R}$ \\
$\hat{\pi}$: pseudo-target density function with support $\mathcal{S}_{\hat{\pi}} \subseteq \mathbb{R}$ such that $\mathcal{S} \subseteq \mathcal{S}_{\hat{\pi}}$ \\
$\widehat{\Pi}$: pseudo-target cumulative distribution function (CDF, continuous) \\
$\widehat{\Pi}^{-1}$: pseudo-target inverse CDF function \\
$x_0$: current state

\Ensure \\ 
$x_1$: new state \\
$u_1$: new state on the transformed scale (which is useful for diagnostics)

\medskip
\hrule
\medskip

\State Draw $v \sim \Unifdist(0, \, h(x_0))$ where $h(x_0) = g(x_0) / \hat{\pi}(x_0)$ \Comment{Define the slice region.}
\State $u_0 \gets \widehat{\Pi}(x_0)$ \Comment{Transform to probability space.}
\State $L \gets 0$, $R \gets 1$
\Loop
\State Draw $u_1 \sim \Unifdist(L, \, R)$ \Comment{Draw candidate.} 
\State $x_1 \gets \widehat{\Pi}^{-1}(u_1)$ \Comment{Back transform to original state space.}
\If{$h(x_1) > v$} \textbf{break} \Comment{If candidate is in slice region, accept.}
\Else 
\If{$u_1 < u_0$} $L \gets u_1$ \Comment{Otherwise, perform shrinkage step.}
\Else~$R \gets u_1$
\EndIf
\EndIf
\EndLoop
\end{algorithmic}
\end{algorithm}

\textit{Quantile slice sampling} proceeds as follows. 
Given an unnormalized target density $g(\theta)$ and a user-specified pseudo-target density $\hat{\pi}(\theta)$, with accompanying  (continuous and computable) CDF $\widehat{\Pi}$ and inverse CDF $\widehat{\Pi}^{-1}$, 
transform to $\psi = \widehat{\Pi}(\theta)$. 
Then, apply uniform simple slice sampling with the shrinkage procedure to 
\begin{align}
    \label{eq:transformedApprox_target}
    \pi_\psi(\psi) \propto h( \widehat{\Pi}^{-1}(\psi) ) \, \Unifdist(\psi; 0, 1) \, ,    
\end{align}
and back transform $\theta^{(s)} \leftarrow \widehat{\Pi}^{-1}(\psi^{(s)})$. 
Samples $\{\theta^{(s)}\}$ from a converged chain have stationary distribution $\Pi$ with density $\pi(\theta)$. 
The procedure for a single update is outlined in Algorithm \ref{alg:quantile_slice}. 
In practice, densities are evaluated on the logarithmic scale for the sake of numerical stability. 
An equivalent procedure employs a simple slice sampler without transformation by drawing $V \mid \theta^{(s-1)} \sim \Unifdist(0, \, h(\theta^{(s-1)}))$ followed by a draw for $\theta^{(s)} \mid V = v$ using Algorithm \ref{alg:generalShrink} with $Q = \widehat{\Pi}$, $x_0 = \theta^{(s-1)}$, and $A = A_h(v)$. 
Although this alternate view is more concise, its practical implementation is generally no more efficient than Algorithm \ref{alg:quantile_slice}, which we find instructive. 
Furthermore, retaining samples $\{ \psi^{(s)} \}$ is useful for evaluating pseudo-target quality. 
Note that a different pseudo-target $\hat\pi$ can be chosen at each iteration of MCMC. 
This is useful if the target full conditional changes with the state, as is the case in Gibbs samplers.

The quality of the $\hat{\pi}$ approximation to $\pi$ affects the efficiency of the algorithm in several ways. 
Well-chosen approximations make the transformed target $h( \widehat{\Pi}^{-1}(\psi) )$ more concave and flat, enlarging the slice region relative to the initial interval and leading to earlier acceptance of proposed draws (as in Figure \ref{fig:intro_cartoon}). 
If the transformed target is less skewed than the original target, the MCMC chain will also exhibit less autocorrelation \citep{planas2018symmetry}. 
A perfect approximation $\hat{\pi} = \pi$ reduces the problem to slice sampling a uniform distribution, wherein the first draw is always accepted. 
Poorly chosen approximations may increase the kurtosis of $h( \widehat{\Pi}^{-1}(\psi) )$ relative to the original target, and therefore decrease efficiency. 
If the target has significant probability mass in regions of the support with low density in the pseudo-target, $h( \widehat{\Pi}^{-1}(\psi) )$ can become unbounded, thereby forfeiting  ergodicity guarantees, or become multimodal, degrading efficiency of the shinkage procedure. 
Furthermore, numerical instabilities involving floating point precision may arise (for example, CDF evaluations round to 1 for all inputs beyond a certain limit). 
We discuss strategies for selecting the pseudo-target in Section \ref{sec:pseudo_target_selection}. 

\begin{figure}[t!]
    \centering
    \includegraphics[width=6.0in]{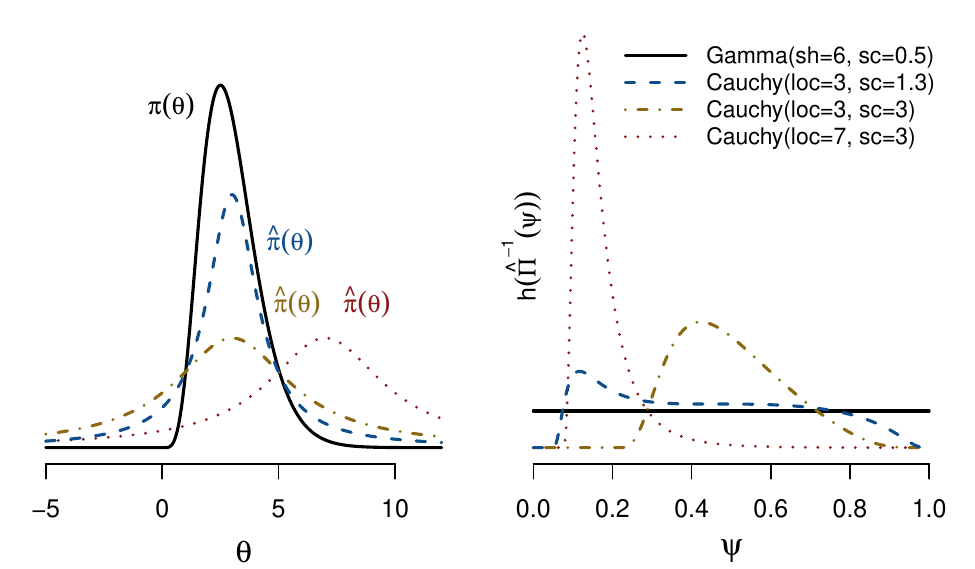}
    \caption{Illustration of transformation to various pseudo-targets. The solid density on the left panel is the target, shown together with three possible pseudo-target densities with varying quality of approximation. The right panel depicts the density to be slice sampled in the transformed space for each of the pseudo-targets. If the approximation is perfect (solid curve), the transformed density is uniform and the first draw is always accepted.}
    \label{fig:transformEffect}
\end{figure}

Figure \ref{fig:transformEffect} illustrates the effect of transformation under different pseudo-targets. 
The solid density on the left panel is the target (a gamma density in this case) overlaid with three approximating location-scale Cauchy pseudo-targets. 
Slice sampling proceeds on the transformed densities in the right panel. 
Aside from larger support, the pseudo-target with location 3 and scale 1.3 approximates the target well; it usually requires fewer than two shrinking steps, and often accepts the first proposal.

\subsection{Relation to independence Metropolis-Hastings}
\label{sec:IMH}

Pseudo-targets appear in independence Metropolis-Hastings (IMH) samplers as proposal distributions. 
\citet{mira2002sliceEfficiency} compare the transition operators of IMH and slice sampling. 
Assuming the target $\pi(\theta) \propto \mathcal{L}(\theta)f(\theta)$, they specify a simple slice sampler that draws directly from truncated $f$ (that is, no hybrid extension), and an IMH sampler that draws proposals from $f$ and evaluates $\mathcal{L}(\theta)$. 
If we introduce pseudo-target $\hat{\pi}(\theta)$, both samplers retain their original structures, substituting $f$ with $\hat{\pi}$ and $\mathcal{L}$ with $h(\theta) = \mathcal{L}(\theta)f(\theta)/\hat{\pi}(\theta)$ from \eqref{eq:pseudoprior}. 
The IMH sampler proposes from $\hat{\pi}(\theta)$ and evaluates $h(\theta)$. 
\citet{mira2002sliceEfficiency} prove (in their Theorem 3) that for any IMH sampler with proposal $\hat{\pi}$, the corresponding slice sampler achieves ``smaller asymptotic variance of sample path averages (on a sweep by sweep basis)." 
Note that the shrinkage procedure is \textit{not} used in this comparison, which assumes that a draw on the slice region can be immediately obtained. 
\citet{latuszynski2024hybrid} and \citet{power2024hybridslice} consider the effect of adding hybrid steps on convergence. 

To build intuition for Mira and Tierney's \citeyearpar{mira2002sliceEfficiency} result, consider the following. The probability that IMH accepts a candidate $\theta^*$ drawn from $\hat{\pi}(\theta)$ is the minimum of 1 and $h(\theta^*) / h(\theta^{(s-1)})$, which is equivalent to accepting the candidate if a variable drawn uniformly on $(0, \, h(\theta^{(s-1)})$ is less than $h(\theta^*)$. 
The slice sampler accepts initial candidate $\theta^*$ if $v < h(\theta^*)$ with intermediate value $v$ drawn uniformly on $(0, \, h(\theta^{(s-1)}))$. 
The two algorithms differ in that rejection in IMH leads to setting $\theta^{(s)} \leftarrow \theta^{(s-1)}$, whereas the slice sampler ensures movement, either by sampling directly on the slice region or by repeatedly drawing from $\widehat{\Pi}$ until a draw within the slice region is found. 

Framing the quantile slice sampler in terms of IMH also suggests one way to evaluate the quality of approximation by the pseudo-target. 
The probability of making a transition in IMH from point $\theta$ is equal to
\begin{align}
    \label{eq:IMHtransProb}
    \int_{\mathcal{S}} \alpha_{h}(\theta, \theta^*) {\widehat{\Pi}(\diff \theta^*)} = \E_{\widehat{\Pi}}( \alpha_{h}(\theta, \theta^*) ) \in [0,1] \, ,
\end{align}
where $\alpha_{h}(\theta, \theta^*) = \min\{ \left( h(\theta^*) / h(\theta) \right), \, 1 \}$, $h(\cdot)$ is defined in \eqref{eq:pseudoprior}, and $\mathcal{S}$ is the support of $\widehat{\Pi}$. Equation \ref{eq:IMHtransProb} can be interpreted as the expected probability of accepting a proposal drawn from $\widehat{\Pi}$ when the current state is $\theta$. 
A successful pseudo-target yields $h(\cdot)$ close to a constant function and $\alpha_{h}(\theta, \theta^*)$ near 1 across $\mathcal{S}$. 

%% file: sec_selecting_pseudo_target.tex
\section{Selecting a pseudo-target}
\label{sec:pseudo_target_selection}

Aided by the transformation underlying \eqref{eq:transformedApprox_target}, we quantify pseudo-target fidelity,
leading to methods for initial or intermediate selection of approximate targets, as well as diagnostic metrics for post-sampling evaluation. 
Although the performance of samplers employing pseudo-targets hinges on the quality of approximation, pairing pseudo-targets with slice sampling and the shrinkage procedure offers a degree of robustness in that even crude approximations can result in reasonably performant samplers. 

General strategies for pseudo-target specification may include Laplace approximations or augmented Gibbs samplers \citep{nishihara2014parallellipslice}, or optimizing metrics for distributions, such as Kullback Leibler divergence \citep{cabezas2023transport}. 
If an approximation to the moments of the target distribution is inexpensive, this could be used to obtain an easy ``moment matching" pseudo-target. 

It is usually important to select a pseudo-target with heavier tails than the target to help ensure boundedness of the importance ratio $h$ in \eqref{eq:pseudoprior}.  
As a general recommendation, we propose selecting within a class of pseudo-target distribution families, in most cases Student-$t$. 
We offer two optimization criteria for selecting parameters within the chosen pseudo-target family in Sections \ref{sec:ESW} and \ref{sec:AUC}. 
The first connects to the IMH acceptance probability. 
The second is less computationally expensive and works well in practice.

\subsection{Quantifying pseudo-target fidelity}
\label{sec:ESW}

We aim to measure approximation quality through its effect on two aspects of algorithm performance: i) autodependence in the Markov chain and ii) cost (in CPU time) of repeated target evaluations in search of the slice region. 
Simple slice samplers with known slice regions are concerned only with the first aspect. 
\citet{planas2018symmetry} observed that autocorrelation from slice samplers is reduced for target distributions with low skewness, and is equal to 0 for centrally symmetric targets. 

\begin{figure}[b!]
    \centering
    \includegraphics[width = 5.0in]{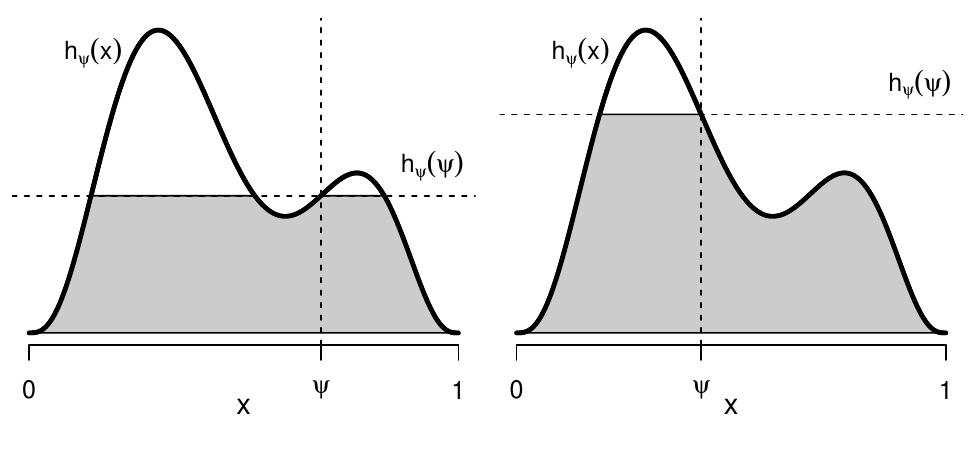}
    \caption{Depiction of a geometric interpretation of $\alpha_h(\psi)$ at two values of $\psi$ for the same generic unnormalized density $h_\psi(\cdot)$. The normalized area $\alpha_h(\psi)$ is calculated as the area of the shaded region divided by $c = \int_0^1 h_\psi(\psi) \diff \psi$.}
    \label{fig:avgSliceWidth_cartoon}
\end{figure}

Repeated target evaluation in search of the slice region is of more practical than theoretical concern. 
The transformation in \eqref{eq:transformedApprox_target} clarifies the problem. 
The procedure of drawing uniformly from an interval suggests that the number of rejections is minimized when the total length of the slice region is maximized. 
Let $h_\psi(\psi) \defeq h(\widehat{\Pi}^{-1}(\psi))$ denote the unnormalized, transformed target density from \eqref{eq:transformedApprox_target} and let $c = \int_0^1 h_\psi(\psi) \diff \psi$. 
We assume that $\sup_\psi h_\psi(\psi) < + \infty$. 
Further denote $W = w(V) \in [0,1]$ as the total length of the slice region as a function of the auxiliary latent variable $V$ drawn uniformly on $(0, \, h_\psi(\psi))$. 
Calculating total length $w(v)$ for $V=v$ may include aggregating lengths of disconnected component intervals of the slice region. 
The expected total slice length, with respect to the marginal (stationary) distribution of $\psi$, is given as 
\begin{align}
    \label{eq:avg_sliceWidth}
    \begin{split}
        \E_h[\E(W \mid \psi)] &= \int_0^1 \left[ \int_{0}^{h_\psi(\psi)} w(v) \frac{1}{h_\psi(\psi)} \diff v \right] \frac{h_\psi(\psi)}{c} \diff \psi \\
        &= \int_0^1 \frac{1}{c} \left[ \int_{0}^{h_\psi(\psi)} w(v) \diff v \right] \diff \psi \\
        &= \int_0^1 \alpha_h(\psi) \diff \psi \, ,        
    \end{split}    
\end{align}
where $\alpha_h(\psi) \defeq c^{-1} \int_{0}^{h_\psi(\psi)} w(v) \diff v$
is interpretable as the normalized area of the region 
$\{ (x, y) : 0 < y < \min(h_\psi(x), \, h_\psi(\psi)), \, 0 < x < 1 \}$,
as depicted in Figure \ref{fig:avgSliceWidth_cartoon} for two values of $\psi$ with generic $h_\psi(\cdot)$. 
Note that 
$\alpha_h(z) = 1$ for $z = \argmax_{x} h_\psi(x)$ and 
that $\alpha_h(\psi)$ is averaged with respect to a \textit{uniform} distribution on $\psi$ rather than the target distribution. 

The transformation $\psi = \widehat{\Pi}(\theta)$ reveals another important connection with IMH. 
In Section \ref{sec:Appendix_sliceWidthProof} of the appendix, we formally state and prove Proposition \ref{prop:sliceWidth}, which establishes equivalence among: i) the expected IMH acceptance probability in \eqref{eq:IMHtransProb}, ii) the expected probability of the slice region under $\widehat{\Pi}$, and iii) the expected total slice width in \eqref{eq:avg_sliceWidth}.

\subsection{An alternate objective function}
\label{sec:AUC}

Now consider scaling $h_\psi(\psi)$ by $m = \max_x(h_\psi(x))$ so that $\underline{h}_\psi(\psi) \defeq h_\psi(\psi) / m$ is enclosed within the unit square. 
We substitute the problem of finding a pseudo-target and corresponding $h_\psi$ that maximizes \eqref{eq:avg_sliceWidth} with the significantly easier problem of maximizing the area under the $\underline{h}_\psi(\psi)$ curve. 
We denote this alternate objective as AUC, formally defined as $\int_0^1 \underline{h}_\psi(\psi) \diff \psi$, which can take on values in $[0,1]$. 

Unfortunately, AUC and \eqref{eq:avg_sliceWidth} do not have a one-to-one correspondence. 
It is possible to shift mass to increase AUC while decreasing \eqref{eq:avg_sliceWidth}. However, both increase as $h_\psi$ approaches a constant function. 
In our experience, pseudo-targets maximizing AUC typically yield high values of \eqref{eq:avg_sliceWidth}. 
Empirical evidence suggests that additional penalties for skewness and for local modes do not yield justifiable gains in pseudo-target quality.

In practice, we recommend maximizing AUC within a family of candidate pseudo-target distributions. 
This can be done prior to sampling via numerical integration of $\underline{h}_\psi(\psi)$. 
Alternatively, AUC can be approximated with a histogram using samples $\{\theta^{(s)}\}$ transformed to $\{\psi^{(s)}\}$ under a proposed pseudo-target. 
Both approaches are demonstrated in Section \ref{sec:illustrations}.

Samples of $\{\psi^{(s)}\}$ can further be used for interpretable sampler diagnostics and tuning. 
The shape of the histogram indicates which adjustments to the pseudo-target can improve efficiency. 
For example, off-center (from 0.5) histograms indicate a location bias. 
Narrow histograms indicate the pseudo-target is too diffuse, while U-shaped histograms indicate the pseudo-target is too narrow and/or has insufficient tail mass. 

%% file: sec_illustrations.tex
\section{Illustrations and sampler performance}
\label{sec:illustrations}

This section examines performance of our quantile slice sampler (QSlice) relative to other popular, easily implemented MCMC samplers. 
We first compare sampler performance on three standard targets in Section \ref{sec:compare_methods}. 
In Section \ref{sec:gprior}, we illustrate use of QSlice in a Bayesian modeling context, as a step within a Gibbs sampler.

In both studies, competing algorithms include random-walk Metropolis (RWM), 
independence Metropolis-Hastings (IMH), 
a uniform slice sampler using 
Neal's \citeyearpar{neal2003slice} shrinkage procedure preceded by the stepping-out procedure (step \& shrink), 
the generalized elliptical slice sampler \citep{nishihara2014parallellipslice}, 
and the latent slice sampler \citep{li2023latent}. 
Performance is evaluated using the effective number of samples (computed with the \texttt{coda} package; \citealp{coda_package}) per CPU second (ESpS). 
Additional details are provided in Section \ref{sec:appendix_sim} of the appendix. 
Implementations of all slice samplers and IMH use the \texttt{qslice} package \citep{qslice_package}.

\input{sec_sim_standard_targets.tex}

\input{sec_gprior.tex}

%% file: sec_sim_standard_targets.tex
\subsection{Simulating standard targets}
\label{sec:compare_methods}

This experiment tests all samplers in isolation on three common families: standard normal, gamma (shape 2.5), and inverse gamma (shape 2). 
The gamma target exhibits skew and light tails, and the inverse-gamma target exhibits extreme skew and heavy tails. 
Competing samplers were tuned for performance in ESpS using preliminary runs. 

Quantile slice and IMH samplers all use Student-$t$ pseudo-targets with left truncation at 0 for targets with positive support. 
Despite differences in the lower quantiles, truncated $t$ distributions adequately approximate skewness and polynomial tails necessary for the inverse-gamma target. 
We employed four strategies select pseudo-target parameters (location, scale, and degrees of freedom in $\{1,\, 5, \, 20\}$): 
i) numerically maximize either mean slice width (MSW; see Section \ref{sec:ESW}) or area under the transformed density curve $\underline{h}_\psi(\psi)$ (AUC; see Section \ref{sec:AUC}); 
ii) maximize histogram-estimated MSW or AUC using 1,000 independent samples from the target; 
iii) Laplace approximation with a Cauchy pseudo-target; 
and iv) crude ``moment" (location and scale) matching between target and a Cauchy pseudo-target (MM). 
We also included a misspecified pseudo-target for which the scale found by the AUC method was inflated by a factor of 4 (AUC--diffuse). 

\begin{figure}[tb]
    \centering
    \includegraphics[width=6.5in]{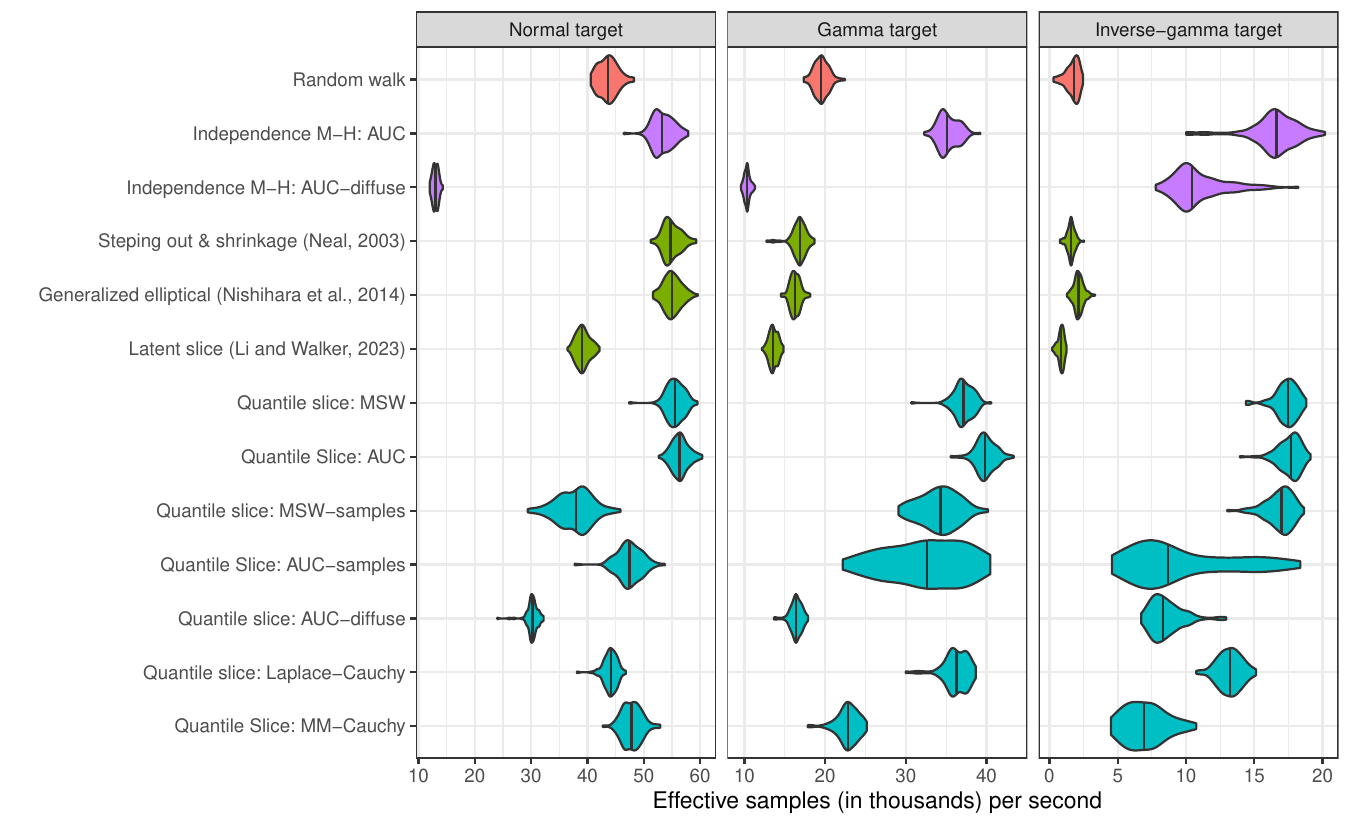}
    \caption{Violin plots summarizing effective samples per CPU second for 100 independent MCMC chains of various samplers on three standard targets. Higher values indicate superior performance. Samplers are grouped by algorithm family and pseudo-target specification method. Vertical lines indicate median effective samples per second.}
    \label{fig:sim_standard_targets}
\end{figure}

Figure \ref{fig:sim_standard_targets} summarizes ESpS from 100 independent MCMC chains using the tuned algorithms. 
We first observe that methods employing pseudo-targets can better react to skewness and heavy tails than RWM and existing slice methods. 
They are, however, sensitive to the quality of approximation; IMH ranges from among the best (with AUC) to worst performance (with AUC--diffuse) when using normal and gamma targets. 
With good approximation, the quick accept/reject decision of IMH can be more efficient than QSlice. 
However, guaranteed moves within the slice region mitigate the consequences of poor approximation. 
QSlice can yield strong performance with mediocre approximations (MM--Cauchy). 
The AUC alternative for pseudo-target selection appears to yield superior performance to the computationally expensive MSW method. 
Despite introducing variability, using initial samples from the target to approximate the target works well. 

%% file: sec_gprior.tex
\subsection{Gibbs sampling: hyper-g prior}
\label{sec:gprior}

Our second illustration uses Zellner's \citeyearpar{zellner1986g} $g$ prior for Bayesian linear regression. 
In this setting, we observe performance of an algorithm embedded within a Gibbs sampler and test sampler diagnostics. 
We also compare two strategies for pseudo-target specification: approximating the full conditional and marginal posterior distributions.

The model is given as
\begin{align}
    \label{eq:g_regression}
    \begin{split}
            \bm{y} \mid \bm{\beta}, \sigma^2 &\sim \mathcal{N}(X\bm{\beta}, \, \sigma^{2} I) \, , \\
    \bm{\beta} \mid \sigma^2, \gamma &\sim \mathcal{N}(\bm{0}, \, \gamma\, \sigma^2 (X^\top X)^{-1}) \, , \quad
    \sigma^{2} \sim \pi_{\sigma^2} \,, \quad
    \gamma \sim \pi_\gamma \, ,
    \end{split}
\end{align}
where $\bm{y}$ is a vector of $n$ (centered/scaled) responses, $X$ is a full-rank $n$ by $p$ matrix of (centered/scaled) covariates, $\bm{\beta}$ is a vector of $p$ regression coefficients, scalar $\sigma^2 > 0$ is the error variance, and scalar $\gamma > 0$ shrinks $\bm{\beta}$ toward $\hat{\bm{\beta}} = (X^\top X)^{-1} X^\top  \bm{y}$. 
The model is completed with an inverse-gamma prior for $\sigma^2$ and the hyper-$g$ prior $\pi_\gamma(\gamma) \propto (1+\gamma)^{a/2}$ proposed by \citet{liang2008mixgprior}, which we adopt after restricting $\gamma < 3p^2$. 

We use the \textit{Motor Trend} road test (\texttt{mtcars}) data from the R \texttt{datasets} package \citep{henderson1981mtcars, Rteam} with $n = 32$ observations of miles per gallon (\texttt{mpg}) as the response and all $p=10$ remaining variables as covariates. 
Our prior specification uses $a = 3$, following \citet{liang2008mixgprior}, and inverse gamma for $\sigma^2$ with shape $5/2$ and scale $5(0.4)^2/2 = 0.4$. 

To sample from the joint posterior distribution of $\{\bm{\beta}, \sigma^2, \gamma\}$, our Gibbs sampler cycles through conjugate updates of $\bm{\beta}$ and $\sigma^2$, and employs one of the competing algorithms to sample from the full conditional of $\gamma$. 
A burn-in phase of 10,000 iterations using step \& shrink for $\gamma$ is followed by an adaptive tuning phase specific to each sampler, 
after which the sampler proceeds for 50,000 timed iterations. 
We timed 100 independent chains for each sampler type. Additional details are given in Section \ref{sec:appendix_gprior} of the appendix. 

\begin{figure}[b!]
    \centering
    \includegraphics[height = 1.85in, trim = 0 0 3 0, clip]{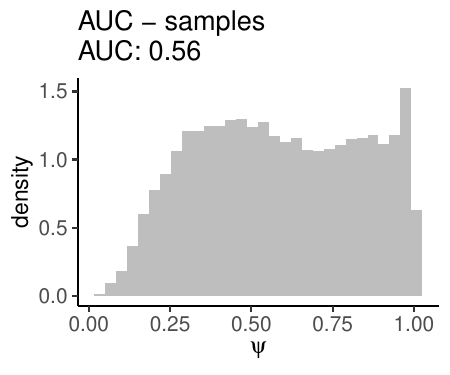}\includegraphics[height = 1.85in, trim = 16 0 3 0, clip]{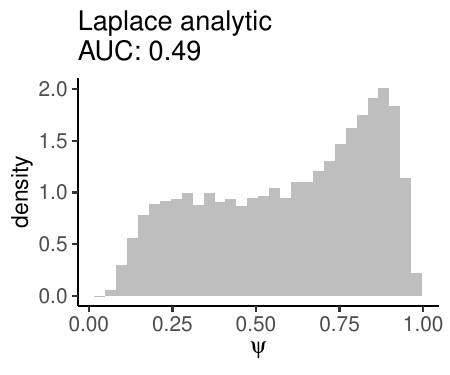}\includegraphics[height = 1.85in, trim = 16 0 0 0, clip]{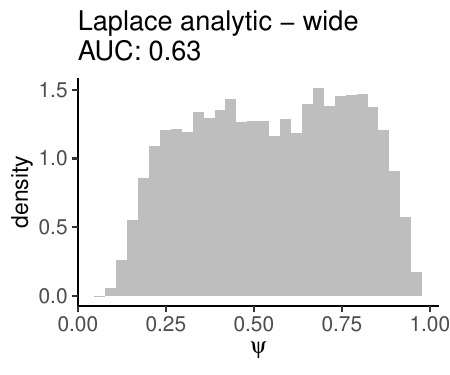}
    \caption{Histograms of samples of $\psi$ from the quantile slice sampler for $\gamma$ under three pseudo-target specification strategies: maximizing AUC over burn-in samples (left), analytic Laplace approximation at each iteration (center), and Laplace approximation with 50\% scale increase (right). The sample-based AUC metric is also reported.}
    \label{fig:psi_hist}
\end{figure}

We employ two strategies for pseudo-target specification, used with the IMH and the generalized elliptical and quantile slice samplers. 
The first constructs the pseudo-target from 2,000 of the burn-in samples of $\gamma$ by finding a Student-$t$ distribution that minimizes the (histogram-based) AUC, thus relying on the marginal posterior distribution of $\gamma$ to approximate the full conditional. 
This approach can be sensitive to dependence of $\gamma$ on $\bm{\beta}$ and $\sigma^2$ in the joint posterior. 
We check for this by monitoring the number of shrinking steps needed to accept a draw at each iteration. 
Another diagnostic uses the sampled $\{ \psi^{(s)} \}$. 
The left panel of Figure \ref{fig:psi_hist} gives a histogram of these samples for a typical run. 

Our second strategy for specifying the pseudo-target is to approximate the full conditional at each iteration. 
The full conditional for $\gamma$ under the hyper-$g$ prior is close to an inverse-gamma distribution, suggesting this family for a pseudo-target. 
Unfortunately, extremely low density values near the origin make inverse gamma a numerically unstable choice for pseudo-targets generally. 
We instead select a truncated Student-$t$, relying on an analytical Laplace approximation (see Section \ref{sec:appendix_gprior} of the appendix for details). 
Although numerical Laplace approximation is feasible, we find it unreliable and slow, requiring respecification at each MCMC iteration. 
One could alternatively set the location and scale of the pseudo-target to ``match" the approximating inverse gamma. 

The histogram of $\{ \psi^{(s)} \}$ in the center panel of Figure \ref{fig:psi_hist} reveals that substantial probability mass in the full conditional lies outside the bulk of the Laplace-approximated $t$ pseudo-target. 
We therefore increase (tune) the scale of the pseudo-target by 50\%, which we refer to as the Laplace--wide pseudo-target (right panel of Figure \ref{fig:psi_hist}). 
This results in a distribution closer to uniform and higher sample AUC values. 

\begin{figure}[tb]
    \centering
    \hspace{-1.15cm}
    \includegraphics[scale = 0.85]{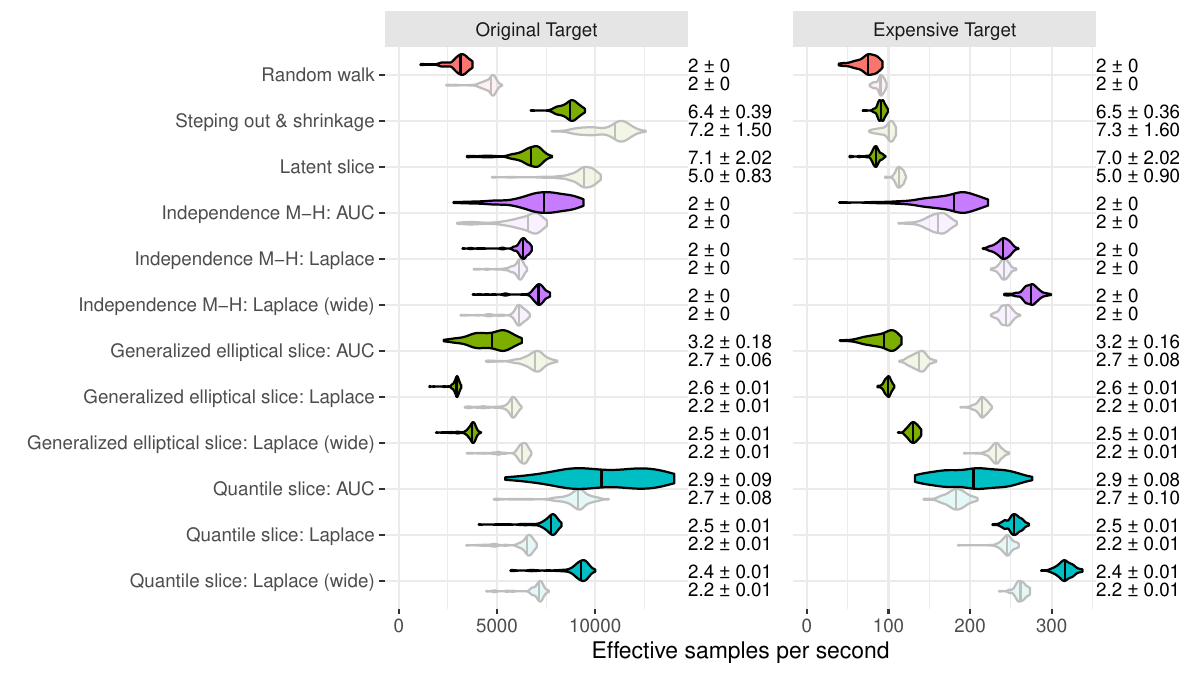}
    \caption{Violin plots summarizing effective samples per CPU second for 100 independent chains of various samplers for $\gamma$ in the $g$-prior example with the original target full conditional (left panel) and a full conditional with superfluous matrix calculations (right panel).
    Higher values indicate superior performance. 
    Numbers on the right report average number of target evaluations per iteration, plus/minus one standard deviation. 
    Samplers are grouped by algorithm family and pseudo-target specification method. 
    Lightly shaded densities show performance of each sampler on the target for $\log(\gamma)$. 
    }
    \label{fig:ESpS_gprior}
\end{figure}

Timing results are summarized with violin plots in Figure \ref{fig:ESpS_gprior}. 
Because it is common to transform variables with restricted support, we also report timing results for each of the samplers applied to sampling $\log(\gamma)$ (shown in light shading). 
This transformation reduces skew in the target and substantially improves performance of several samplers.

The step \& shrink algorithm sampling $\log(\gamma)$ outperforms QSlice tuned with initial samples, on average. 
However, 
step \& shrink also averages more than twice as many target evaluations as QSlice, which becomes a liability in scenarios with computationally expensive targets. 
This is illustrated in the right panel of Figure \ref{fig:ESpS_gprior}; in these runs, the full-conditional target for $\gamma$ was artificially burdened with superfluous matrix calculations common in regression modeling. 
As the target expense increases, IMH (which always evaluates the target only twice) and Qslice become the preferred methods (for ESpS) in this scenario with high-fidelity pseudo-targets. 
Despite volatility from dependence on a preliminary MCMC sample, Qslice with AUC--samples yields excellent performance, even within a Gibbs sampler. 

%% file: sec_multivariate.tex
\section{Multivariate sampling}
\label{sec:multivariate}

\citet{neal2003slice} explores multivariate slice-sampling schemes with a single latent variable to avoid augmentation in each dimension. 
Notably, the shrinkage procedure trivially generalizes to shrink along each axis of a hyperrectangle; an algorithm is given as Figure 8 in \citet{neal2003slice}. 
While we recommend the generalized elliptical slice sampler \citep{nishihara2014parallellipslice} for most multivariate targets with unrestricted support, a multivariate extension of the quantile slice sampler can be useful in cases with nonstandard or restricted support and where a natural approximation to the target exists. See Section \ref{sec:DHR} for an example. 

The factorization in \eqref{eq:pseudoprior} applies readily to random vectors $\bm{\theta} \defeq (\theta_1, \ldots, \theta_D)$ with $\pi(\bm{\theta}) \propto h(\bm{\theta}) \hat{\pi}(\bm{\theta})$ and $h(\bm{\theta}) \defeq g(\bm{\theta}) / \hat{\pi}(\bm{\theta})$. 
In Sections \ref{sec:multi_indep} and \ref{sec:pseu-condseq}, we identify two general approaches to construct bijective mappings from $\bm{\theta}$ to $\bm{\psi} \in (0,1)^D$ that yield multivariate quantile slice samplers. 
The methodology is not restricted to these two cases. 
Because we always transform to $\bm{\psi}$ supported on a subset of the unit hypercube $(0,1)^D$, we provide a theoretical result in this scenario only. 
A proof is given in Section \ref{sec:multivariate_unif_ergod} of the appendix.

\begin{proposition}
    \label{prop:multivariate_unif_ergodicity}
    Assume target density $\pi(\bm{\psi}) \propto \tilde{h}(\bm{\psi}) I\{ \bm{\psi} \in (0,1)^D \}$ with lower- \\ semicontinuous function $\tilde{h}: (0,1)^D \mapsto [0, \infty)$. 
    Consider a simple slice sampler that defines the slice region with $V \mid \bm{\psi} \sim \Unifdist(0, \tilde{h}(\bm{\psi}))$ and uses the shrinkage procedure in Figure 8 of \citet{neal2003slice}, initialized on the unit hypercube, as a hybrid step targeting $p(\bm{\psi} \mid V=v) \propto I\{ \bm{\psi} \in A_{\tilde{h}}(v) \}$ with $A_{\tilde{h}}(v) \defeq \{ \bm{\psi} : v < \tilde{h}(\bm{\psi}) \}$. 
    The transition kernel for this hybrid slice sampler is $\pi$-reversible. 
    If $\sup_{\bm{\psi}} \tilde{h}(\bm{\psi}) < \infty$, the resulting Markov chain is uniformly ergodic. 
\end{proposition}

\subsection{Uncorrelated targets}
\label{sec:multi_indep}

If elements of $\bm{\theta}$ are approximately independent in the target distribution (or undergo a decorrelating rotation; \citealp{tibbits2014factorslice}), then a pseudo-target built from independent components may suffice. 
That is, we can use $\hat{\pi}(\bm{\theta}) = \hat{\pi}_1(\theta_1) \times \cdots \times \hat{\pi}_D(\theta_D)$ and transform to $\bm{\psi} = (\widehat{\Pi}_1(\theta_1), \ldots, \widehat{\Pi}_D(\theta_D)) \in (0,1)^D$
with 
univariate CDFs corresponding to the densities $\{ \hat{\pi}_d \}$. 
The sampler proceeds as described in Proposition \ref{prop:multivariate_unif_ergodicity} using target $\pi_{\bm{\psi}}(\bm{\psi}) \propto \tilde{h}(\bm{\psi}) \prod_d \Unifdist(\psi_d ; 0, 1)$ with $\tilde{h}(\bm{\psi}) = h( (\widehat{\Pi}_1^{-1}(\psi_1), \ldots, \widehat{\Pi}_D^{-1}(\psi_D)))$. 

Natural transformations to approximately independent targets exist in some common cases. 
For example, if $\bm{\theta}$ is compositional (that is, nonnegative entries that sum to one) and approximately satisfies complete neutrality, we can exploit the stick-breaking representation of the generalized Dirichlet distribution of \citet{connor1969}. 
This representation uses $D-1$ mutually independent beta-distributed variables to construct $\bm{\theta}$.

\subsection{Approximations from cascading conditional distributions}
\label{sec:pseu-condseq}

Chained conditional distributions provide another avenue for mapping multivariate random vectors to $(0,1)^D$. 
If a decomposition $\hat{\pi}_1(\theta_1)\, \hat{\pi}_2(\theta_2 \mid \theta_1) \cdots \hat{\pi}_D(\theta_D \mid \theta_1, \ldots, \theta_{D-1})$ of $\hat{\pi}(\bm{\theta})$ is readily available, we can transform to $\bm{\psi}$ using the cascading sequence of conditional CDFs, that is, $\bm{\psi} = ( \widehat{\Pi}_1(\theta_1), \, \widehat{\Pi}_2(\theta_2 \mid \theta_1), \, \ldots, \, \widehat{\Pi}_D(\theta_D \mid \theta_1, \ldots, \theta_{D-1}) ) \in (0,1)^D$. 
Conveniently, the Jacobian of this transformation has determinant equal to \\ $\left[ \hat{\pi}_1( \widehat{\Pi}_1^{-1}(\psi_1) ) \times \cdots \times \hat{\pi}_D( \widehat{\Pi}_D^{-1}(\psi_D \mid \psi_1, \ldots, \psi_{D-1}) ) \right]^{-1}$, again admitting a uniform slice sampler with a multivariate shrinkage procedure applied to $\pi_{\bm{\psi}}(\bm{\psi}) \propto h( (\widehat{\Pi}_1^{-1}(\psi_1), \, \ldots, \, \widehat{\Pi}_D^{-1}(\psi_D \mid \psi_1, \ldots, \psi_{D-1}))) \prod_d \Unifdist(\psi_d ; 0, 1)$. 
We use this construction in Section \ref{sec:DHR}.

Unlike univariate quantile slice samplers and multivariate samplers using independent pseudo-target components in Section \ref{sec:multi_indep}, the uniform slice sampler with shrinkage applied to $\bm{\psi}$ with cascading conditional pseudo-targets is not equivalent to slice sampling $\bm{\theta}$ with the generalized shrinkage procedure using the same pseudo-targets. 
This is because shrinking hyperrectangles in the Cartesian space of $\bm{\theta}$ does not map to hyperrectangles on the space of $\bm{\psi}$. 
Although both approaches are valid, they are distinct. 

%% file: sec_DHR.tex
\section{Multivariate illustration}
\label{sec:DHR}

\begin{figure}[tb]
    \centering
    \includegraphics[width=6.4in]{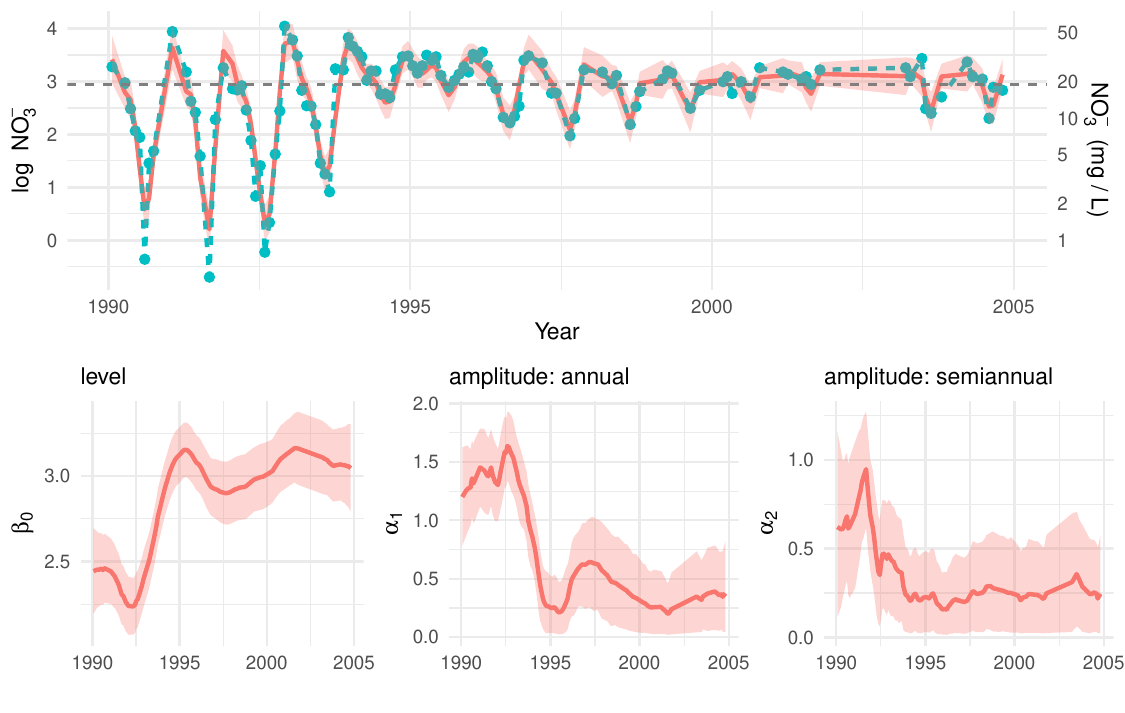}
    \caption{Top panel -- Time series of NO$_3^-$ concentrations for a stream in France (dashed with points) with fit and uncertainty in the mean (solid), and target concentration (at 19 mg/L; \citealp{abbott2018trends}). Bottom panel -- Time-varying parameter estimates, with uncertainty, for the process level ($\beta_0$, left), amplitude of the annual cycle ($\alpha_1$, center), and amplitude of the semiannual cycle ($\alpha_2$, right) of log-NO$_3^-$.}
    \label{fig:river}
\end{figure}

We demonstrate a multivariate quantile slice sampler with an application targeting conditional block updates of constrained, time-varying parameters (TVPs) within a non-Gaussian dynamic linear model (DLM). 
We fit DLMs to time series of nitrate (NO$_3^-$) concentrations in streams in France, collected at approximately monthly intervals \citep{NaiadesData}. 
Nitrate pollution, typically attributed to agricultural activity, can lead to harmful algal blooms, prompting management efforts that rely on monitoring long-term trends and seasonality \citep{abbott2018trends}. 
The top panel of Figure \ref{fig:river} shows a time series for a stream that exhibits drift in both. 

\subsection{Dynamic harmonic regression model}
\label{sec:DHR_model}

We fit a dynamic harmonic regression (DHR) model that combines 
seasonality with TVPs \citep{young1999DHR}. 
DHR is traditionally parameterized using Fourier coefficients, conflating variability in the amplitudes and phases of the cycles and thus complicating prior elicitation. 
We model amplitudes and phases directly, which introduces constraints to the model and in MCMC. 

Let $y_t$ represent the log-NO$_3^-$ concentration at time $s_t$, measured continuously in years, for observation index $t = 1, \ldots, T$; here, $T = 117$. The DHR model is given as
\begin{align}
    \label{eq:DHR_obs}
    y_t = \beta_{0,t} + \alpha_{1,t} \cos\left(2\pi[s_t + \phi_1]\right) + \alpha_{2,t} \cos\left(2\pi[2s_t + \phi_2]\right) + \epsilon_t \, ,
\end{align}
with time-indexed intercepts $\{\beta_{0,t}\}$, amplitudes of the annual cycle $\{\alpha_{1,t}\}$ with static phase $\phi_1 \in (0, 1)$, amplitudes of the semiannual cycle $\{\alpha_{2,t}\}$ with static phase $\phi_2 \in (0, 1)$, and independent observation noise $\epsilon_t \simiid \mathcal{N}(0, \sigma^2)$. 
The TVPs are given initial distributions and evolve in the prior according to random walks with continuous-time adjustment. 
For example, $\alpha_{1,0} \sim \mathcal{N}(0, 4^2) \, 1\{\alpha_{1,0} \ge 0\}$ and $\alpha_{1,t} \mid \alpha_{1,t-1} \sim \mathcal{N}(\alpha_{1,t-1}, \, 0.1875[s_t - s_{t-1}]) \, 1\{\alpha_{1,t} \ge 0\}$ with $1\{\cdot\}$ indicating truncation to enforce positivity constraints. 
The model is completed with priors on $\phi_1$, $\phi_2$, and $\sigma^2$. 
Further details are given in Section \ref{sec:appendix_DHR} of the appendix.

The lower panel of Figure \ref{fig:river} summarizes posterior inferences for the TVPs with evolving means and 90\% credible intervals. 
Posterior distributions for the amplitude parameters press against the lower boundary of 0, which helps with identifiability in \eqref{eq:DHR_obs} since negative amplitude values shift the corresponding phases. 
Such behavior is sometimes observed in samples from the standard forward-filter and backward-sampling (FFBS; \citealp{carter1994ffbs, fruhwirth1994ffbs}) update for TVPs with no boundary constraints.

Despite nonidentifiability, FFBS is useful for posterior sampling if used as the basis of a pseudo-target for the full conditional distributions of $\bm{\alpha}_j \defeq (\alpha_{j,1}, \ldots, \alpha_{j,T})$, $j=1,2$. 
Specifically, the algorithm exploits the Markovian prior and conditional independence to decompose the joint full conditional for $\bm{\alpha}_j$ into cascading univariate Gaussian densities, provided the initial state, evolution, and observation distributions are linear and Gaussian. 
From these we construct a joint pseudo-target for $\bm{\alpha}_j$ following the procedure in Section \ref{sec:pseu-condseq}, using truncated Gaussian or Student-$t$ components with location and scale parameters obtained from the corresponding FFBS distributions. Section \ref{sec:appendix_DHR} of the appendix outlines the algorithm to update $\balpha_j$.

\subsection{Sampler comparison}
\label{sec:DHR_comparison}

We compare the efficiency of our multivariate quantile slice (MQSlice) sampler against other candidate algorithms for the amplitude updates. 
We notably exclude the elliptical slice sampler, for which truncation is problematic, and gradient-assisted MCMC, which is not easily incorporated within the Gibbs sampler.

The first generic competitor is IMH with the FFBS-based pseudo-target described above as a proposal distribution, similar to the strategies discussed in \citet{fearnhead2011mcmchandbook}. 
We also compare against multivariate versions of two slice samplers: 
i) with the standard shrinkage procedure on a hyperrectangle with no pseudo-target and no transformation (MSlice; \citealp{neal2003slice}), and ii) the latent slice sampler of \citet{li2023latent}. 
Both slice samplers restrict the hyperrectangle to exclude negative amplitudes. 

We add two comparisons that are specific to state-space modeling. 
First is the FFBS algorithm, which is not a true competitor because it does not address truncation and samples a nonidentified model. 
FFBS is the only algorithm that updates $\{ \{\beta_{0,t}\}, \bm{\alpha}_1, \bm{\alpha}_2 \}$ as a single block, and it is included as a benchmark. 
The second specialized algorithm is a particle MCMC sampler that uses conditional sequential Monte Carlo within a Gibbs framework. 
Our implementation specifically uses the bootstrap filter with a backward-sampling step \citep[Ch.~16]{chopin2020book}. 

Table \ref{tab:DHRriver} reports results from the full simulation study, which consisted of 50 replicate timing runs for each algorithm, following the protocol outlined in Section \ref{sec:appendix_sim} of the appendix. 
Reported effective sample sizes aggregate from two chains and summarize (mean, minimum) across the $3T = 351$ parameters in $\{ \{\beta_{0,t}\}, \bm{\alpha}_1, \bm{\alpha}_2 \}$. 
We also report summaries of the number of target evaluations per iteration and the potential scale reduction factors (PSRF, commonly known as R-hat; \citealp{gelman1992MCMC}) for which values below 1.05 are commonly accepted as indicating convergence.

\begin{table}
    \centering
    \begin{tabular}{llrrrr}
    \toprule
        Sampler & Settings & PSRF & ESpS & ESpS & Target \\
        &  &  & (mean) & (minimum) & evaluations \\
    \midrule
       IMH & pseudo: Gaussian & $1.184 \pm 0.905$ & $88.0 \pm 3.8$ & $28.88 \pm 4.77$ & $2$ (fixed) \\
           & pseudo: $t(\text{df}=5)$ & $1.594 \pm 1.316$ & $4.7 \pm 1.3$ & $0.58 \pm 0.16$ & $2$ (fixed) \\
       MSlice & $w = 1.5$ & $1.052 \pm 0.014$ & $3.5 \pm 0.2$ & $0.23 \pm 0.02$ & $12.45 \pm 0.01 $ \\
       Latent & $r = 3.33$ & $1.051 \pm 0.012$ & $3.9 \pm 0.2$ & $0.27 \pm 0.03$ & $10.52 \pm 0.01 $ \\
       MQSlice & pseudo: Gaussian & $1.0006 \pm 0.0001$ & $79.6 \pm 2.4$ & $41.63 \pm 1.76$ & $3.37 \pm 0.01 $  \\
           & pseudo: $t(\text{df}=5)$ & $1.0032 \pm 0.0006$ & $4.3 \pm 0.2$ & $1.23 \pm 0.06$ & $6.83 \pm 0.01 $ \\
       Particle & $25+1$ particles & $1.0003 \pm 0.0001$ & $53.4 \pm 1.8$ & $25.35 \pm 1.12$ & n/a \\
       FFBS & & $1.0002 \pm 0.0001$ & $85.4 \pm 2.9$ & $65.58 \pm 2.74$ & n/a \\
    \bottomrule
    \end{tabular}
    \caption{Performance of various samplers for the time-varying parameters in the dynamic harmonic regression model. Summaries give mean plus/minus one standard deviation across 50 independent runs of each algorithm. PSRF for each run is the upper bound of a 95\% confidence interval for the Gelman-Rubin diagnostic, averaged for all time-varying parameters. 
    ESpS in each run refers to the mean (or minimum) effective samples per CPU second of all time-varying parameters. 
    Target evaluations are reported per iteration and averaged across both amplitude parameter vectors.}
    \label{tab:DHRriver}
\end{table}

MQSlice is the only highly performing slice sampler in this study, achieving strong ESpS and stable diagnostics with high-fidelity Gaussian pseudo-targets (which require care, as the importance ratio $h$ can become unbounded), and is still the preferred slice sampler when using a more robust Student-$t$ pseudo-target. 
PSRF results indicate that the general slice samplers struggle with parameter vectors of length $T=117$. 
The high number of target evaluations reveals that these samplers take so many shrinkage steps at each iteration that the space is not efficiently explored, though they yield a reasonable fit. 
Our results confirm that particle MCMC is well suited to posterior sampling for non-Gaussian TVPs. 

High ESpS for IMH reveals that FFBS-based truncated Gaussian pseudo-targets faithfully approximate the joint full conditionals in this example. 
IMH outperforms quantile slice sampling in mean ESpS, although IMH mixing performance is somewhat volatile. 
Performance of MQSlice and IMH is sensitive to the tails of high-dimensional pseudo-targets. 

This timing experiment employed custom implementations of each competing algorithm, written in R (in which several functions call compiled C code). It is important to recognize that benchmarking results will vary by hardware, operating system, programming language, and implementation. For example, the MQSlice sampler written for DHR runs approximately 3-8 times faster than when implemented with a more general framework for pseudo-target specification in the \texttt{qslice} package \citep{qslice_package}.

%% file: sec_discussion.tex
\section{Discussion}
\label{sec:discussion}

This paper revisits slice sampling with the shrinkage procedure of \citet{neal2003slice}, extending its applicability. 
The probability integral transform allows shrinking from an automatic and universal starting point, and enables introduction of pseudo-targets, which can reduce the number of shrinking steps. 
The resulting quantile slice sampler can be viewed as a slice-sampling analog of independence Metropolis Hastings. 

We have found the stepping-out and shrinkage procedure of \citet{neal2003slice} to be effective as a general-purpose sampler for univariate targets (with low skewness) when evaluation is inexpensive. 
Its performance is often insensitive to specification of the tuning parameter. 
Absent a natural candidate to approximate a univariate target, we recommend employing stepping-out and shrinkage during a burn-in phase and using the initial samples to develop a pseudo-target for use with a quantile slice sampler. 

As noted in Section \ref{sec:multivariate} and demonstrated in Section \ref{sec:DHR}, quantile slice samplers can be used to leverage an available approximation to a complex or constrained multivariate target. 
We generally recommend elliptical slice sampling \citep{nishihara2014parallellipslice} for multivariate targets with unrestricted support, as its use of shrinkage on only one dimension is usually more efficient than shrinking along each axis. 

A primary limitation of quantile slice sampling is its sensitivity to the quality of target approximation. 
If one is concerned with optimal sampling speed, a high-fidelity pseudo-target is required. 
We recommend using pseudo-targets that bound the importance ratio to ensure uniform ergodicity. 
Badly misspecified pseudo-targets can result in numerical instability. 
Nevertheless, inadequate pseudo-targets can quickly be identified with a preliminary sample, and even crude approximations can result in reasonable performance.

Our standard unimodal pseudo-targets can work with multimodal targets as long as the pseudo-target sufficiently straddles all modes. 
The shrinkage procedure will nevertheless elevate autocorrelation in this scenario. 
We note that mixture densities are not effective pseudo-targets because evaluation of the inverse-CDF of a mixture distribution requires costly numerical root-finding. 

Quantile and elliptical slice sampling belong to a broader class that could be called importance slice samplers with generalized shrinkage procedures. 
Novel shrinkage procedures could give rise to samplers that outperform Metropolis-Hastings counterparts. 
Areas with potential include targeted shrinkage and shrinkage procedures for multivariate discrete variables with support too large to enumerate, which can benefit from improved balance between local and global exploration.

%% file: appx_proofs_transform.tex
\section{Proofs of Propositions in Section \ref{sec:transform}}
\label{sec:Appendix_proofTransform}

We present three lemmas that establish equivalence of the Markov chains in the original and transformed spaces. 
They facilitate the proofs of Propositions \ref{prop:generalShrink} and \ref{prop:shrink_in_slice} that follow. 
Most results assume the following conditions.

\begin{conditions}
    \label{cond:assumptions}
Let $\theta$ be a real-valued random variable with target distribution $\Pi$ having density $\pi(\theta) \propto \mathcal{L}(\theta) f(\theta)$, factored into 
\begin{enumerate}
    \item probability density $f(\theta)$ with support on Lebesgue-measurable $\mathcal{S}_F \subseteq \mathbb{R}$, and 
    \item nonnegative, lower semicontinuous function $\mathcal{L}(\theta)$ with support on Lebesgue-measurable $\mathcal{S} \subseteq \mathcal{S}_F$. 
\end{enumerate}
The support of $\pi$ is then $\mathcal{S} \cap \mathcal{S}_F = \mathcal{S}$. 
Let $F(\theta)$ denote the distribution function associated with $f$ and assume that its inverse $F^{-1}(u)$ can be computed for all $u \in (0,1)$.   
\end{conditions}

\begin{definition}
\label{def:simpleslice}
Let $K_{\theta}(\theta_0, \cdot)$ denote the transition kernel for the simple slice sampler that alternates drawing $(V \mid \theta) \sim \Unifdist(0,\mathcal{L}(\theta))$ and $p(\theta \mid V = v) \propto f(\theta) I\{ \theta : v < \mathcal{L}(\theta) \}$, and let $K_{\phi}(\phi_0, \cdot)$ denote the transition kernel for the uniform slice sampler targeting $\Pi_\phi$ under transformation $\phi = F(\theta)$ with density $\pi_\phi(\phi) \propto \mathcal{L}(F^{-1}(\phi))$.     
\end{definition}

\subsection{Equivalence of slice samplers under transformation}
\label{sec:equiv_chains}

\begin{lemma}
    \label{lemma:equiv_chains}
    Under Conditions \ref{cond:assumptions}, transition kernels $K_\theta$ and $K_\phi$ are equivalent. 
    That is, for any $\theta_0 \in \mathcal{S}$ and $B \in \mathcal{B}(\mathcal{S})$ (the Borel sets of $\mathcal{S}$), we have $K_{\theta}(\theta_0, B) = K_{\phi}(\phi_0, \img_F(B))$, where $\phi_0 = F(\theta_0)$ and $\img_F(B) \in \mathcal{B}(\img_F(\mathcal{S}))$ with $\img_F(\mathcal{S}) \subseteq (0,1)$. 
    Thus, the Markov chain $\{\phi^{(t)} : t = 1, \ldots \}$ resulting from $K_\phi$ is equivalent to the Markov chain $\{\theta^{(t)}\}$ resulting from $K_\theta$. 
\end{lemma}
\noindent \textbf{Proof}: 
Let $\theta_0 \in \mathcal{S}$ and $\phi_0 = F(\theta_0) \in \img_F(\mathcal{S}) \subseteq (0,1)$. 
Integrating over the slice-defining variable, $V$, the density for the transition kernel of the simple slice sampler is
\begin{align}
    \label{eq:transkern_density}
    k_\theta(\theta_1 \mid \theta_0) = \int_0^{\mathcal{L}(\theta_0)} \frac{1}{\mathcal{L}(\theta_0)} \frac{f(\theta_1) I\{\theta_1 \in A_\mathcal{L}(v) \}}{F(A_\mathcal{L}(v))} \diff v \, ,
\end{align}
where $A_\mathcal{L}(v) = \{ \theta : v < \mathcal{L}(\theta) \}$ can be expressed as a countable union of open sets in $\mathcal{S}$ by the lower semicontinuity of $\mathcal{L}$, and $F(A) \defeq \Pr(\theta \in A)$ under distribution $F$ for any $A \in \mathcal{B}(\mathcal{S}_F)$. 
Changing variables, let $\phi_1 = F(\theta_1) \in \img_F(\mathcal{S})$ so that $\theta_1 = F^{-1}(\phi_1)$ with derivative $f(F^{-1}(\phi_1))^{-1}$. 
Also, for any $v > 0$, it holds that 
\begin{align*}
\img_F(A_\mathcal{L}(v)) &= \img_F(\{ \theta : v < \mathcal{L}(\theta) \}) \\ 
&= \{ \phi : v < \mathcal{L}(F^{-1}(\phi)) \} \\ 
&= A_{\mathcal{L} \circ F^{-1}}(v)
\end{align*}
and 
\begin{align*}
F(A_\mathcal{L}(v)) &= \int_{A_\mathcal{L}(v)} f(\theta) \diff \theta \\
&= \int_{A_{\mathcal{L} \circ F^{-1}}(v)} \frac{ f(F^{-1}(\phi)) } { f(F^{-1}(\phi)) } \diff \phi \\
&= \lvert A_{\mathcal{L} \circ F^{-1}}(v) \rvert
\end{align*}
where $\lvert A \rvert$ denotes the Lebesgue measure of set $A$. 
The density in \eqref{eq:transkern_density} under transformation thus becomes 
\begin{align}
    \label{eq:transkern_density_unif}
    k_\phi(\phi_1 \mid \phi_0) &= \int_0^{\mathcal{L}(F^{-1}(\phi_0))} \frac{1}{\mathcal{L}(F^{-1}(\phi_0))} \frac{f(F^{-1}(\phi_1)) I\{F^{-1}(\phi_1) \in A_\mathcal{L}(v) \}}{ f(F^{-1}(\phi_1)) F(A_\mathcal{L}(v))} \diff v \nonumber \\
    &= \int_0^{\mathcal{L}(F^{-1}(\phi_0))} \frac{1}{\mathcal{L}(F^{-1}(\phi_0))} \frac{ I\{ \phi_1 \in A_{\mathcal{L} \circ F^{-1}}(v) \} }{ \lvert A_{\mathcal{L} \circ F^{-1}}(v) \rvert } \diff v \, .
\end{align}
This is precisely the density for the transition kernel of a uniform slice sampler targeting a density proportional to $\mathcal{L} \circ F^{-1}(\cdot)$. 
Integrating \eqref{eq:transkern_density} and \eqref{eq:transkern_density_unif} over $B$ and $\img_F(B)$, respectively, for any $B \in \mathcal{B}(\mathcal{S})$ yields the desired result. 
\hfill $\square$

\subsection{Equivalence of shrinkage procedures under transformation}
\label{sec:equiv_shrinkage}

As with the simple slice sampler, the shrinkage procedure in Algorithm \ref{alg:generalShrink} preserves equivalence of Markov kernels under bijective transformation. 
We first provide a definition for the shrinking intervals in terms of a sequence of points that will correspond with rejected values. 

\begin{definition}
\label{def:shrinking_intervals}
Let $\{ z_j : j = 0, 1, \ldots\}$ denote a countable sequence with $z_0 \in \mathcal{S} \subseteq \mathbb{R}$, and $z_j \in \mathcal{I}_j(z_{0:(j-1)}) \defeq (L_{j}, R_{j})$, for $j \ge 1$, with $z_{0:{(j-1)}} \defeq (z_0, z_1, \ldots, z_{j-1})$, and $L_j$ and $R_j$ defined recursively as follows. 
Let $L_1 = \inf \mathcal{S}$ and $R_1 = \sup \mathcal{S}$. 
Then, $(L_j, R_j) =(z_{j-1}, R_{j-1})$ if $z_{j-1} < z_0$ and $(L_j, R_j) = (L_{j-1}, z_{j-1})$ otherwise, for $j > 1$. 
\end{definition}

Let $K_{Q,A}$ denote the transition kernel associated with Algorithm \ref{alg:generalShrink}, repeated below. 

\medskip

\begin{algorithm}[h!]
\caption*{\textbf{Algorithm \ref{alg:generalShrink}} Generalized shrinkage procedure (repeated from Section \ref{sec:transform}).}
\begin{algorithmic}
\Require \\
$Q$: continuous distribution to be sampled with unrestricted support $\mathcal{S}_Q$ \\
$A$: subset of $\mathcal{S}_Q$ to which the support of $x$ is to be restricted \\
$x_0$: current state (in $A$)

\Ensure \\ 
$x_1$: new state \\

\medskip
\hrule

\State $L \gets \inf \mathcal{S}_Q$, $R \gets \sup \mathcal{S}_Q$
\Loop
\State Draw $x_1 \sim Q(x \mid L \le x \le R)$ \Comment{Sample $Q$ restricted to $x \in (L, R)$.}
\If{$ x_1 \in A$} \textbf{break}
\Else
    \If{$x_1 < x_0$} $L \gets x_1$
    \Else~$R \gets x_1$
    \EndIf
\EndIf
\EndLoop

\end{algorithmic}
\end{algorithm}

\begin{lemma}
    \label{lemma:equiv_shrinkage}
    Let $Q$ be a continuous and monotonic distribution function on Lebesgue-measurable support $\mathcal{S}_Q \subseteq \mathbb{R}$ and let $A \subseteq \mathcal{S}_Q$ with $Q(A) \defeq \Pr_Q(X \in A) > 0$. 
    Let $x_0 \in A $ 
    and $u_0 = Q(x_0) \in \img_Q(A) \subseteq(0,1)$. 
    Then the transition kernels $K_{Q,A}(x_0, \cdot)$ and $K_{\Unifdist(0,1), \img_Q(A)}(u_0, \cdot)$ are equivalent. 
    That is, for any $x_0 \in A$ 
    and $B \in \mathcal{B}(\mathcal{S}_Q)$ (the Borel sets of $\mathcal{S}_Q$), we have $K_{Q,A}(x_0, B) = K_{\Unifdist(0,1), \, \img_Q(A)}(u_0, \img_Q(B))$ where $\img_Q(B) \in \mathcal{B}((0,1))$. 
\end{lemma}

\noindent \textbf{Proof}: 
First note that by the probability integral transform, if $X \sim Q$, then $U = Q(X) \sim \Unifdist(0,1)$ and $Q(B) = \Pr_Q(X \in B) = \Pr(U \in \img_Q(B)) = \lvert \img_Q(B) \rvert$, where $\lvert E \rvert$ denotes the Lebesgue measure of $E \subseteq \mathbb{R}$.
Also, let $q(x) = \diff Q(x) / \diff x$ denote the density of $Q$.

Under Algorithm \ref{alg:generalShrink}, 
we have $K_{Q,A}(x_0, B) = \Pr(x_1 \in A \cap B \mid x_0)$ since $\Pr(x_1 \in B \setminus A \mid x_0) = 0$, where $B \setminus A$ is the intersection of $B$ and the complement of $A$, denoted $A^C$. 
Then, $K_{Q,A}(x_0, B) = \Pr(x_1 \in A \cap B \mid x_0) = \sum_{i=1}^\infty \Pr(x_1 \in A \cap B \ \text{on $i$th step} \mid x_0)$ since if $x_1 \in A \cap B$ at any step, it is immediately accepted. 
The result follows if we can show that the shrinkage procedure preserves $\Pr(x_1 \in A \cap B \ \text{on $i$th step} \mid x_0) = \Pr(u_1 \in \img_Q(A \cap B) \ \text{on $i$th step} \mid u_0)$, where $u_1 = Q(x_1)$, for all $i \in \mathbb{N}^+$.

We proceed with an inductive argument. 
When $i = 1$, we have $\Pr(x_1 \in A \cap B \ \text{on $1$st step} \mid x_0 ) = Q(A \cap B)$. 
Under transformation, this becomes $\Pr(x_1 \in A \cap B \ \text{on $1$st step} \mid x_0 ) = \Pr(u_1 \in \img_Q(A \cap B) \ \text{on $1$st step} \mid u_0 ) = \lvert \img_Q(A \cap B) \rvert$. 

Let $f_s(z_1, \ldots, z_j \mid z_0)$ denote the joint density of a sequence of random variables drawn from distribution $Q$ with the shrinking mechanism of Algorithm \ref{alg:generalShrink}, but without rejection. 
Using Definition \ref{def:shrinking_intervals}, the joint density is given as 
\begin{align}
    \label{eq:shrink_jointdens}
    f_s(z_1, z_2, \ldots, z_J \mid z_0) 
    &= \prod_{j=1}^J \frac{q(z_j) I\{ z_j \in \mathcal{I}_j(z_{0:(j-1)}) \}}{Q(\mathcal{I}_j(z_{0:(j-1)}))} \, ,
\end{align}
where $\mathcal{I}_1(z_0) = (\inf \mathcal{S}_Q, \, \sup \mathcal{S}_Q)$ and $Q(\mathcal{I}_1(z_0)) = 1$.
Letting $z_0 = x_0$, we can integrate this density over the desired set to obtain 
\begin{align}
    \label{eq:shrink_jointprob}
    \Pr(x_1 &\in A \cap B \ \text{on $i$th step} \mid x_0) = \Pr(z_1 \notin A, z_2 \notin A, \ldots, z_{i-1} \notin A, z_i \in A \cap B \mid z_0) \nonumber \\ 
    &= \int_\mathbb{R} \cdots \int_\mathbb{R} f_s(z_1, z_2, \ldots, z_i \mid z_0) \left[ \prod_{j = 1}^{i-1} I\{ z_j \notin A \} \right]  I\{z_i \in A \cap B\}  \diff z_1 \cdots \diff z_i \nonumber \\
    &= \int_\mathbb{R} \cdots \int_\mathbb{R} \left[ \prod_{j=1}^{i-1} \frac{q(z_j) I\{ z_j \in \mathcal{I}_j(z_{0:(j-1)}) \setminus A \}}{Q(\mathcal{I}_j(z_{0:(j-1)}))} \right] \frac{q(z_i) I\{ z_i \in \mathcal{I}_{i}(z_{0:(i-1)}) \cap A \cap B \}}{Q(\mathcal{I}_i(z_{0:(i-1)}))} \diff z_1 \cdots \diff z_i \, \nonumber \\
    &= \int_\mathbb{R} \cdots \int_\mathbb{R} \left[ \prod_{j=1}^i q(z_j) \right] \left[ \prod_{j=1}^{i-1} \frac{I\{ z_j \in \mathcal{I}_j(z_{0:(j-1)}) \setminus A \}}{Q(\mathcal{I}_j(z_{0:(j-1)}))} \right] \frac{ I\{ z_i \in \mathcal{I}_{i}(z_{0:(i-1)}) \cap A \cap B \}}{Q(\mathcal{I}_i(z_{0:(i-1)}))} \diff z_1 \cdots \diff z_i \,.
\end{align}

Consider the transformation $\tilde{z}_1 = Q(z_1), \ldots, \tilde{z}_i = Q(z_i)$ with inverse transformation $z_j = Q^{-1}(\tilde z_j)$ for $j = 1, \ldots, i$, and Jacobian $\prod_{j = 1}^i [q(Q^{-1}(\tilde z_j))]^{-1}$. 
Because $Q$ is continuous and increasing, it is straightforward to show the following: 
\begin{align*}
\img_Q(\mathcal{I}_j(z_{0:(j-1)}) \cap A^C) &= \img_Q(\mathcal{I}_j(z_{0:(j-1)})) \cap \img_Q(A^C) \, , \\
\img_Q(\mathcal{I}_j(z_{0:(j-1)}) \cap A \cap B) &= \img_Q(\mathcal{I}_j(z_{0:(j-1)})) \cap \img_Q(A \cap B) \, ,
\end{align*}
for any $j \ge 1$. 
It is similarly straightforward to show that $\img_Q(A^C) = [\img_Q(A)]^C$. 

Also, the sequence $\{ \img_Q ( \mathcal{I}_j(z_{0:(j-1)}) ) : j = 1, \ldots \}$ is identical to the sequence $\{  \tilde{\mathcal{I}}_j(\tilde z_{0:(j-1)}) ) : j = 1, \ldots \}$ obtained by applying the definition of shrinking intervals to the sequence $\{ \tilde z_j = Q(z_j) : j \ge 0 \}$, for any $z_0 \in \mathcal{S}_Q, z_1 \in \mathcal{S}_Q, \ldots$. 
To see this, let $\tilde z_0 = Q(z_0)$, $\tilde L_1 = 0$, $\tilde R_1 = 1$, and $\tilde{\mathcal{I}}_1(\tilde z_0) = (0,1) = \img_Q \mathcal{I}_1(z_0)$. 
Then because $Q$ is continuous and increasing, $\tilde z_j > \tilde z_{0}$ if and only if $z_j > z_{0}$, for all $j \ge 1$. 
Thus, the shrinking step that yields $\tilde{\mathcal{I}}_2(\tilde z_{0:1})$ will shrink on the same side of $\tilde z_1$ (right or left) that $\mathcal{I}_2(z_{0:1})$ will shrink in reference to $z_0$. 
That is, $\tilde L_2$ equals $Q(L_2)$ if $L_2 > \inf \mathcal{S}_Q$ and 0 otherwise, and $\tilde R_2$ equals $Q(R_2)$ if $R_2 < \sup \mathcal{S}_Q$ and 1 otherwise. 
Thus $\tilde{\mathcal{I}}_2(\tilde z_{0:1}) = \img_Q(\mathcal{I}_2(z_{0:1}))$. 
Now assume that $\tilde{\mathcal{I}}_j(\tilde z_{0:(j-1)}) = \img_Q(\mathcal{I}_j(z_{0:(j-1)}))$ for some $j > 2$. 
Then, introducing $\tilde z_j = Q(z_j)$ will again result in a shrinking step that shrinks on the same side of $\tilde z_0$ as $z_0$, ensuring that $\tilde L_{j+1} = Q(L_{j+1})$ if $L_{j+1} > \inf \mathcal{S}_Q$ and 0 otherwise, and $\tilde R_{j+1} = Q(R_{j+1})$ if $R_{j+1} < \sup \mathcal{S}_Q$ and 1 otherwise. 
Thus $\tilde{\mathcal{I}}_{j+1}(\tilde z_{0:j}) = \img_Q(\mathcal{I}_{j+1}(z_{0:j}))$ and, by induction, the sequences $\{ \img_Q ( \mathcal{I}_j(z_{0:(j-1)}) ) : j = 1, \ldots \}$ and $\{  \tilde{\mathcal{I}}_j(\tilde z_{0:(j-1)}) ) : j = 1, \ldots \}$ are identical. 

Under the transformation, the probability in \eqref{eq:shrink_jointprob} becomes
\begin{align}
    \label{eq:shrink_jointprob_tnx}
    \Pr(x_1 &\in A \cap B \ \text{on $i$th step} \mid x_0) \nonumber \\ 
    &= \int_0^1 \cdots \int_0^1 \left[ \prod_{j=1}^i \frac{q(Q^{-1}(\tilde z_j))}{q(Q^{-1}(\tilde z_j))} \right] \left[ \prod_{j=1}^{i-1} \frac{I\{ \tilde z_j \in \img_Q( \mathcal{I}_j(Q^{-1}(\tilde z_{0:(j-1)})) \setminus A) \}}{\lvert \img_Q(\mathcal{I}_j(Q^{-1}(\tilde z_{0:(j-1)}))) \rvert} \right] \times \nonumber \\ 
    & \qquad \qquad \qquad\frac{ I\{ \tilde z_i \in \img_Q (\mathcal{I}_{i}(Q^{-1}(\tilde z_{0:(i-1)})) \cap A \cap B) \}}{\lvert \img_Q(\mathcal{I}_i(Q^{-1}(\tilde z_{0:(i-1)}))) \rvert} \diff \tilde{z}_1 \cdots \diff \tilde{z}_i \nonumber \\
    &= \int_0^1 \cdots \int_0^1 \left[ \prod_{j=1}^{i-1} \frac{I\{ \tilde z_j \in \img_Q( \mathcal{I}_j(Q^{-1}(\tilde z_{0:(j-1)})) ) \cap \img_Q(A^C) \}}{\lvert \img_Q(\mathcal{I}_j(Q^{-1}(\tilde z_{0:(j-1)}))) \rvert} \right] \times \nonumber \\ 
    & \qquad \qquad \qquad\frac{ I\{ \tilde z_i \in \img_Q (\mathcal{I}_{i}(Q^{-1}( \tilde z_{0:(i-1)} ))) \cap \img_Q (A \cap B) \}}{\lvert \img_Q(\mathcal{I}_i(Q^{-1}(\tilde z_{0:(i-1)}))) \rvert} \diff \tilde{z}_1 \cdots \diff \tilde{z}_i \nonumber \\
    &= \int_0^1 \cdots \int_0^1 \left[ \prod_{j=1}^{i-1} \frac{I\{ \tilde z_j \in  \tilde{\mathcal{I}}_j(\tilde z_{0:(j-1)})) \setminus \img_Q(A) \}}{\lvert \tilde{\mathcal{I}}_j(\tilde z_{0:(j-1)}) \rvert} \right] \times \nonumber \\ 
    & \qquad \qquad \qquad\frac{ I\{ \tilde z_i \in \tilde{\mathcal{I}}_{i}(\tilde z_{0:(i-1)}) \cap \img_Q (A \cap B) \}}{\lvert \tilde{\mathcal{I}}_i(\tilde z_{0:(i-1)}) \rvert} \diff \tilde{z}_1 \cdots \diff \tilde{z}_i \, ,
\end{align}
Where $Q^{-1}(\tilde z_{0:j})$ represents $(Q^{-1}(\tilde z_0), \ldots, Q^{-1}(\tilde z_j))$. 
The final line of \eqref{eq:shrink_jointprob_tnx} is exactly the value of $\Pr(u_1 \in \img_Q(A \cap B) \ \text{on $i$th step} \mid u_0)$ that would be obtained from applying Algorithm \ref{alg:generalShrink} with $Q$ uniform on $(0,1)$. 
Thus, we have $\Pr(x_1 \in A \cap B \ \text{on $i$th step} \mid x_0) = \Pr(u_1 \in \img_Q(A \cap B) \ \text{on $i$th step} \mid u_0)$ for all $i \ge 1$ and, consequently, $K_{Q,A}(x_0, B) = K_{\Unifdist(0,1), \, \img_Q(A)}(u_0, \img_Q(B))$.
\hfill $\square$

\subsection{Proof of Proposition \ref{prop:generalShrink}}
\label{sec:proof_generalShrink}

\textbf{Proposition \ref{prop:generalShrink}}: 
\textit{
The Markov transition kernel $K_{Q,A}$ outlined in Algorithm \ref{alg:generalShrink} using $Q$ with continuous and monotonic distribution function on support $\mathcal{S}_Q \subseteq \mathbb{R}$, and $A \subset \mathcal{S}_Q$ with $Q(A) > 0$, is $Q_A$-reversible. If initialized in $A$, the resulting Markov chain is uniformly ergodic on $A$.
}

\noindent \textbf{Proof}: 
To show that the Markov chain resulting from iterated application of $K_{Q,A}$ is uniformly ergodic, we verify Doeblin's condition (Theorem 6.59 in \citealp{robert2004MC}) by establishing that i) the chain is aperiodic, and ii) $A$ is a small set. 
Aperiodicity of the chain follows immediately from positivity:  
assuming the distribution function of $Q$ is continuous and strictly increasing on $\mathcal{S}_Q$, the first step of the transition kernel with $L = \inf \mathcal{S}_Q$ and $R = \sup \mathcal{S}_Q$ ensures that all of $A$ has positive density under $K_{Q,A}$.

Set $C \subseteq \mathcal{S}_Q$ is said to be small if there exist $t \in \{1,2,\ldots\}$ and nonzero measure $\nu_t$ such that for all $x \in C$ and for all $B \in \mathcal{B}(\mathcal{S}_Q)$ (the Borel sets of the sample space, $\mathcal{S}_Q$), we have
\begin{align}
\label{eq:minorizing}
K^t(x, B) \geq \nu_t(B) \, ,
\end{align}
where $K^t$ denotes $t$ applications of transition kernel $K$ (Definition 6.19 in \citealp{robert2004MC}). 
We take 
$\mathcal{S}_Q$ to be the unrestricted support of $Q$, 
$K = K_{Q, A}$, 
$t = 1$, and 
$\nu_1 = Q_A(\cdot) Q(A) = Q(\cdot \cap A)$, which is nonzero for any nonempty measurable subset of $A$. 
Let $x \in A$ and $B \in \mathcal{B}(\mathcal{S}_Q)$. 
Then,
\begin{align}
\label{eq:A_small}
K_{Q,A}^1(x, B) 
&\geq \Pr(\text{first proposed } x_1 \in B \mid x, A) \nonumber \\
&= Q(B) \, \\
&\geq Q(B \cap A) \, \nonumber \\
&= \nu_1(B). \nonumber
\end{align}
The first line takes only the first of countably many mutually exclusive outcomes tracking on which proposal $x_1$ falls in $B$. 
Unlike subsequent proposals that are subject to shrinkage steps, the first proposal comes directly from $Q$ and does not depend on $x$ or $A$. 
Thus the probability that the first proposal falls in $B$ is simply $Q(B)$. 
The minorizing condition \eqref{eq:minorizing} is met, and $A$ is therefore a small set.

We next establish reversibility of $K_{Q,A}$, which is sufficient for $Q_A$ invariance, by showing that it satisfies detailed balance. 
\citet{neal2003slice} showed that $K_{Q,A}$ is reversible in the case that $Q$ is uniform; here, we show it for any $Q$ satisfying the conditions of the proposition. 
Let $B_0 \in \mathcal{B}(\mathcal{S}_Q)$ and $B_1 \in \mathcal{B}(\mathcal{S}_Q)$. 
Detailed balance requires that $\int_{B_0} g_A(x) K_{Q,A}(x, B_1) \diff x = \int_{B_1} g_A(x) K_{Q,A}(x, B_0) \diff x$, where $g_A(x) \defeq q(x) I\{x \in A \} / Q(A)$ denotes the density of the restricted target. 
The transition kernel can be represented as $K_{Q,A}(x,B) = \sum_{i=1}^\infty \Pr(x_1 \in A \cap B \ \text{on $i$th step} \mid x)$, where the summand is given in \eqref{eq:shrink_jointprob}. 
Each of these terms operates on the same sequence $\{ z_j \}$, with corresponding $\{ \mathcal{I}_j(z_{0:(j-1)}) \}$. 
Only one term will be active for any given sequence. 

The key insight for establishing reversibility is that for any sequence $(x, z_1, \ldots, z_{i-1}, z_i)$ yielding positive transition density, the corresponding sequence of shrinking intervals will be \textit{identical} to the sequence of intervals arising from $(z_i, z_1, \ldots, z_{i-1}, x)$. 
This is because all rejected points lie outside the interval $(\min(x, z_i),\, \max(x, z_i))$, which is symmetric in $x$ and $z_i$, causing identical shrinking steps.
Furthermore, for any $(x, z_i)$ pair, both $x$ and $z_i$ are contained in every interval in $\{\mathcal{I}_j(z_{0:(j-1)}) : j = 1, \ldots,i \}$. 
Letting 
\begin{align*}
    r_i(z_{0:(i-1)}) \defeq \prod_{j=1}^{i-1} \frac{q(z_j) I\{ z_j \in \mathcal{I}_j(z_{0:(j-1)}) \setminus A \}}{Q(\mathcal{I}_j(z_{0:(j-1)}))} 
\end{align*}
and $z_0 = x$, we have
\begin{align}
     \int_{B_0} & g_A(x) K_{Q,A}(x, B_1) \diff x  \nonumber \\
     &= \int_{B_0} \frac{q(x)I\{x \in A \}}{Q(A)} \sum_{i=1}^\infty \int_{B_1} \int_\mathbb{R} \cdots \int_\mathbb{R} r_i(z_{0:(i-1)}) \frac{q(z_i) I\{ z_i \in \mathcal{I}_{i}(z_{0:(i-1)}) \cap A  \}}{Q(\mathcal{I}_i(z_{0:(i-1)}))} \diff z_1 \cdots \diff z_i \diff x \, \nonumber \\
     &=  \int_{B_0} \sum_{i=1}^\infty  \int_{B_1} \int_\mathbb{R} \cdots \int_\mathbb{R} r_i(z_{0:(i-1)}) \frac{ I\{ x, z_i \in \mathcal{I}_{i}(z_{0:(i-1)}) \cap A \}}{Q(\mathcal{I}_i(z_{0:(i-1)}))}  \frac{q(x)q(z_i)}{Q(A)} \diff z_1 \cdots \diff z_i \diff x \, \nonumber \\
     &= \sum_{i=1}^\infty  \int_{B_0} \int_{B_1} \int_\mathbb{R} \cdots \int_\mathbb{R} r_i(z_{0:(i-1)}) \frac{ I\{ x, z_i \in \mathcal{I}_{i}(z_{0:(i-1)}) \cap A \}}{Q(\mathcal{I}_i(z_{0:(i-1)}))}  \frac{q(x)q(z_i)}{Q(A)} \diff z_1 \cdots \diff z_i \diff x \, \nonumber \\
     &= \sum_{i=1}^\infty  \int_{B_1} \int_{B_0} \int_\mathbb{R} \cdots \int_\mathbb{R} r_i(z_{0:(i-1)}) \frac{ I\{ x, z_i \in \mathcal{I}_{i}(z_{0:(i-1)}) \cap A \}}{Q(\mathcal{I}_i(z_{0:(i-1)}))}  \frac{q(x)q(z_i)}{Q(A)} \diff z_1 \cdots \diff z_i \diff x \, \nonumber \\
     &=  \int_{B_1} \sum_{i=1}^\infty  \int_{B_0} \int_\mathbb{R} \cdots \int_\mathbb{R} r_i(z_{0:(i-1)}) \frac{ I\{ x, z_i \in \mathcal{I}_{i}(z_{0:(i-1)}) \cap A \}}{Q(\mathcal{I}_i(z_{0:(i-1)}))}  \frac{q(x)q(z_i)}{Q(A)} \diff z_1 \cdots \diff z_i \diff x \, \nonumber \\
     &= \int_{B_1} \frac{q(x)I\{x \in A \}}{Q(A)} \sum_{i=1}^\infty \int_{B_0} \int_\mathbb{R} \cdots \int_\mathbb{R} r_i(z_{0:(i-1)}) \frac{q(z_i) I\{ z_i \in \mathcal{I}_{i}(z_{0:(i-1)}) \cap A  \}}{Q(\mathcal{I}_i(z_{0:(i-1)}))} \diff z_1 \cdots \diff z_i \diff x \, \nonumber \\
     &= \int_{B_1} g_A(x) K_{Q,A}(x, B_0) \diff x \, . \nonumber
\end{align}
\noindent We can interchange the order of summation and integration by Tonelli's theorem. 
Swapping the regions of integration in the central steps is justified because all shrinking intervals are identical and both $x$ and $z_i$ belong to the final interval.
Thus, $K_{Q,A}$ is $Q_A$-reversible.

\hfill $\square$

\subsection{Equivalence of slice samplers with shrinkage}
\label{sec:equiv_chains_shrinkage}

\begin{lemma}
    \label{lemma:equiv_chains_shrinkage}
    Let $\tilde{K}_\theta$ and $\tilde{K}_\phi$ represent the transition kernels from Definition \ref{def:simpleslice}, but with the generalized shrinkage and original (uniform) shirnkage procedures as hybrid steps, respectively.
    Under Conditions \ref{cond:assumptions}, transition kernels $\tilde{K}_\theta$ and $\tilde{K}_\phi$ are equivalent. 
    That is, for any $\theta_0 \in \mathcal{S}$ and $B \in \mathcal{B}(\mathcal{S})$ (the Borel sets of $\mathcal{S}$), we have $\tilde{K}_{\theta}(\theta_0, B) = \tilde{K}_{\phi}(\phi_0, \img_F(B))$, where $\phi_0 = F(\theta_0) \in \img_F(\mathcal{S}) \subseteq (0,1)$ and $\img_F(B) \in \mathcal{B}(\img_F(\mathcal{S}))$. 
    Thus, the Markov chain $\{\phi^{(t)} : t = 1, \ldots \}$ resulting from $\tilde{K}_\phi$ is equivalent to the Markov chain $\{\theta^{(t)}\}$ resulting from $\tilde{K}_\theta$. 
\end{lemma}

\noindent \textbf{Proof}:
We have
\begin{align}
    \label{eq:transkern_wShrink}
    \tilde{K}_\theta(\theta_0, B) = \int_0^{\mathcal{L}(\theta_0)} \frac{1}{\mathcal{L}(\theta_0)} K_{F, A_\mathcal{L}(v)}(\theta_0, B) \diff v \, ,
\end{align}
where $K_{F, A}$ is the transition kernel associated with Algorithm \ref{alg:generalShrink}. 
Changing variables, \eqref{eq:transkern_wShrink} becomes 
\begin{align}
    \label{eq:transkern_wShrink_transformed}
    \int_0^{\mathcal{L}(F^{-1}(\phi_0))} \frac{1}{\mathcal{L}(F^{-1}(\phi_0))} K_{\Unifdist(0,1), \, A_{\mathcal{L} \circ F^{-1}}(v) }(\phi_0, \, \img_F(B)) \diff v \, ,
\end{align}
by Lemma \ref{lemma:equiv_shrinkage}. 
The transition kernel in \eqref{eq:transkern_wShrink_transformed} is equal to  $\tilde{K}_\phi(\phi_0, \, \img_F(B))$, the transition kernel from Lemma \ref{lemma:equiv_chains} using a uniform shrinkage procedure with initial interval $(0,1)$ on target density proportional to $\mathcal{L} \circ F^{-1}(\cdot)$.
\hfill $\square$

\subsection{Proof of Proposition \ref{prop:shrink_in_slice}}
\label{sec:proof_shrinkinslice}

\textbf{Proposition \ref{prop:shrink_in_slice}}:

\textit{
    Assume target density $\pi(\theta) \propto \mathcal{L}(\theta) f(\theta)$ and Conditions \ref{cond:assumptions}. 
    The simple slice sampler that defines the slice region with $V \mid \theta \sim \Unifdist(0, \mathcal{L}(\theta))$ and implements Algorithm \ref{alg:generalShrink} as a hybrid step targeting $p(\theta \mid V=v) \propto f(\theta) I\{ \theta \in A_\mathcal{L}(v) \}$ is $\pi$-reversible. 
    Furthermore, if $\sup_\theta \mathcal{L}(\theta) < \infty$, then this hybrid simple slice sampler is uniformly ergodic. 
}

\noindent \textbf{Proof}: 
\citet{latuszynski2024hybrid} showed (their Lemma 1) that reversibility of an embedded hybrid step with respect to the uniform distribution over $A_\pi(v)$ for all $v \in (0, \sup_\theta{\pi(\theta)})$ is sufficient for the reversibility of a uniform slice sampler using that hybrid step. 
Reversibility of the uniform shrinkage procedure established by \citet{neal2003slice} and in Proposition \ref{prop:generalShrink} for any measurable $A$ implies reversibility of uniform slice samplers using the shrinkage procedure in Algorithm \ref{alg:generalShrink} (with $Q$ uniform). 
Then, applying this result for transformed $\phi = F(\theta)$ with target $\pi_\phi(\phi) \propto \mathcal{L}(F^{-1}(\phi)) I\{ \phi \in \img_F(\mathcal{S}) \}$ with $\img_F(\mathcal{S}) \subseteq (0,1)$, 
equivalence of the Markov chains $\{\phi^{(s)}\}$ and $\{\theta^{(s)}\}$ from Lemma \ref{lemma:equiv_chains_shrinkage} extends \citeauthor{latuszynski2024hybrid}'s result to establish reversibility of simple slice samplers using the general shrinkage procedure in Algorithm \ref{alg:generalShrink}, subject to the conditions in Proposition \ref{prop:generalShrink}.

Ergodicity depends on the target. 
We next show that if $\mathcal{L}$ is bounded, then uniform slice sampling on target $\mathcal{L}(F^{-1}(\phi))$ with the shrinkage procedure yields a uniformly ergodic Markov chain. 
First, consider Algorithm \ref{alg:generalShrink} with $Q$ uniform on $(0,1)$ and $A = \cup_i A_i$, a possibly countable union of disjoint subintervals of the unit interval. 
Given the current point $\phi_0$, the probability that the final accepted point $\phi_1$ belongs to subinterval $A_i$ is 
\begin{align}
    \label{eq:shrinkage_prob_subinterval}
    \begin{split}
      \Pr(\phi_1 \in A_i \mid \phi_0, A) 
    &\ge \Pr(\text{first proposed} \ \phi_1 \in A_i \mid \phi_0, A) \\
    &= \lvert A_i \rvert
    \end{split}
\end{align}
where $\lvert A_i \rvert$ is the length, or Lebesgue measure of $A_i$. 
The first line takes only the first of countably many mutually exclusive outcomes tracking the proposal on which $\phi_1$ accepted. 
The final equality recognizes that the first proposal is uniform over $(0,1)$. 

We use the probabilities from \eqref{eq:shrinkage_prob_subinterval} to characterize the density of the finally accepted $\phi_1$, given $A$ and $\phi_0$. 
Conditional on $\phi_1 \in A_i$, the density of $\phi_1$ is uniform over $A_i$. 
Therefore,
\begin{align}
    \label{eq:uniform_shrink_density}
    \begin{split}
        p_{\phi_1}(\phi_1 \mid \phi_0, A) &= \sum_i \frac{1}{\lvert A_i \rvert} \Pr(\phi_1 \in A_i \mid \phi_0, A) I\{ \phi_1 \in A_i \} \\
        &\ge \sum_i \frac{\lvert A_i \rvert}{\lvert A_i \rvert} I\{ \phi_1 \in A_i \} \\
        &= \sum_i I\{ \phi_1 \in A_i \} \, ,
    \end{split}
\end{align}
where the inequality uses \eqref{eq:shrinkage_prob_subinterval}. 
Thus, the density of the accepted point is bounded below by 1 everywhere in $A$. 

We can now establish uniform ergodicity of the uniform slice sampler with shrinkage and support on the unit interval (USSU) by verifying Doeblin's condition. 
Specifically, we show that the sampler is aperiodic and that the entire support of $\pi_\phi(\phi)$, or $\img_F(\mathcal{S}) \subseteq (0,1)$, is a small set. 
Following the strategy for Theorem 6 in \citet{mira2002sliceEfficiency}, we write the transition kernel of the slice sampler, $K_\text{USSU}$, integrating over the slice-defining latent variable $V$. 
Note that the sets $A_{\tilde{\mathcal{L}}}(v) = \{ \phi : v < \tilde{\mathcal{L}}(\phi) \}$, where $\tilde{\mathcal{L}}(\phi) \defeq \mathcal{L}(F^{-1}(\phi))$, can be expressed as a countable union of disjoint intervals, as assumed in \eqref{eq:shrinkage_prob_subinterval} and \eqref{eq:uniform_shrink_density}. 
Let $\pi_\phi(\phi) \propto \tilde{\mathcal{L}}(\phi)I\{ \phi \in \img_F(\mathcal{S}) \}$ and $\sup_\phi{\tilde{\mathcal{L}}(\phi)} < \infty$.
Then the single-step transition density from $\phi_0 \in \img_F(\mathcal{S})$ to $\phi_1 \in \img_F(\mathcal{S})$ of $K_\text{USSU}$ is 
\begin{align}
    \label{eq:unif_ergodicity_unif_slice}
    \begin{split}
        k_\text{USSU}(\phi_0, \phi_1) &= \int_0^{\tilde{\mathcal{L}}(\phi_0)} \frac{1}{\tilde{\mathcal{L}}(\phi_0)} p_{\phi_1}(\phi_1 \mid \phi_0, A_{\tilde{\mathcal{L}}}(v)) \diff v \\
        &= \frac{1}{\tilde{\mathcal{L}}(\phi_0)} \int_0^{\min(\tilde{\mathcal{L}}(\phi_0), \, \tilde{\mathcal{L}}(\phi_1))} p_{\phi_1}(\phi_1 \mid \phi_0, A_{\tilde{\mathcal{L}}}(v)) \diff v \\
        &\ge \frac{1}{\tilde{\mathcal{L}}(\phi_0)} \int_0^{\min(\tilde{\mathcal{L}}(\phi_0), \, \tilde{\mathcal{L}}(\phi_1))} 1 \diff v \\
        &= \frac{\min(\tilde{\mathcal{L}}(\phi_0), \, \tilde{\mathcal{L}}(\phi_1))}{\tilde{\mathcal{L}}(\phi_0)} \\
        &\ge \frac{\tilde{\mathcal{L}}(\phi_1)}{\sup_\phi{\tilde{\mathcal{L}}(\phi)}} \\
        &= \frac{c_{\tilde{\mathcal{L}}}}{\sup_\phi{\tilde{\mathcal{L}}(\phi)}} \pi_\phi(\phi_1) \, ,
    \end{split}
\end{align}
where $c_{\tilde{\mathcal{L}}} = \int_0^1 \tilde{\mathcal{L}}(\phi) \diff \phi$. 
Thus $\pi_\phi(\phi)$ itself, which is positive on all of $\img_F(\mathcal{S})$, supplies the minorizing measure.

With $\sup_\phi \tilde{\mathcal{L}}(\phi) = \sup_\theta \mathcal{L}(\theta) < \infty$, Lemma \ref{lemma:equiv_chains_shrinkage} extends uniform ergodicity from $\{ \phi^{(s)} \}$ to the equivalent Markov chain $\{ \theta^{(s)} \}$ with target $\pi(\theta) \propto \mathcal{L}(\theta)f(\theta)$ that uses Algorithm \ref{alg:generalShrink} as a hybrid step for sampling $p(\theta \mid V=v) \propto f(\theta) I\{ \theta \in A_\mathcal{L}(v) \}$.
\hfill $\square$

%% file: appx_proof_sliceWidth.tex
\section{Proof of Proposition \ref{prop:sliceWidth}}
\label{sec:Appendix_sliceWidthProof}

\begin{proposition}
    \label{prop:sliceWidth}
    Let $\pi(\theta)$ and $\hat{\pi}(\theta)$ represent target and pseudo-target densities with respective Lebesgue-measurable supports $\mathcal{S}_\pi \subseteq \mathcal{S}_{\hat{\pi}} \subseteq \mathbb{R}$ and distribution functions $\Pi$ and $\widehat{\Pi}$. Further let $\psi = \widehat{\Pi}(\theta)$, $\pi(\theta) = c_g g(\theta)$ for some normalizing constant $c_g$, $h(\theta) = g(\theta) / \hat{\pi}(\theta)$ on $\mathcal{S}_{\pi}$, and $h_\psi(\psi) = h(\widehat{\Pi}^{-1}(\psi))$. Assume $\sup_\theta h(\theta) < \infty$. Finally, let $\theta^{(s)}$ represent a draw from $\Pi$, with corresponding $\psi^{(s)} = \widehat{\Pi}(\theta^{(s)})$. Consider an IMH sampler using target $g(\theta)$ with proposal $\widehat\Pi$, a simple slice sampler using target $h(\theta)\hat{\pi}(\theta)$, and a uniform simple slice sampler using target $h_\psi(\psi)$. Then the following three quantities are equal to one another: 
    \begin{enumerate}
        \item[Q1.] Expected acceptance probability of IMH proposal, \\ $\E_{\widehat{\Pi}}[ \alpha_{h}(\theta^{(s)}, \theta^*)  \mid \theta^{(s)} ]$ from \eqref{eq:IMHtransProb}, where $\alpha_{h}(\theta^{(s)}, \theta^*) \defeq \min\left( h(\theta^*) / h(\theta^{(s)}), \, 1 \right)$.
        \item[Q2.] Expected probability of the slice region under $\widehat\Pi$ computed as \\ $\E[\widehat{\Pi}(\{ \theta^* : V < h(\theta^*) \}) \mid \theta^{(s)}]$ for $V \sim \Unifdist(0, \, h(\theta^{(s)}))$.
        \item[Q3.] Expected total slice width $\E[W \mid \psi^{(s)}]$ with $W = w(V), \ V \sim \Unifdist(0, \, {h}_\psi(\psi^{(s)}))$.
    \end{enumerate}
    Furthermore, $\E_{\Pi} \big[ \E_{\widehat{\Pi}}( \alpha_{h}(\theta^{(s)}, \theta^*)  ) \, \big] = \int_{\mathcal{S}_\pi} \E_{\widehat{\Pi}}( \alpha_{h}(\theta^{(s)}, \theta^*) ) \Pi(\diff\theta)$ 
    and $\E_h[\E(W \mid \psi^{(s)})]$ from \eqref{eq:avg_sliceWidth} are the same quantity.
\end{proposition}

\noindent \textbf{Proof}: We begin by establishing that quantities Q1 and Q2 are equal. 
Given the current state $\theta^{(s)}$ and proposal $\theta^* \sim \widehat{\Pi}$, the IMH acceptance probability is equal to 1 if $h(\theta^*) \ge h(\theta^{(s)})$ and equal to the ratio $h(\theta^*) / h(\theta^{(s)})$ otherwise. 
To simulate an event (call it ``accept $\theta^*$") that occurs with this probability, draw $V \sim \Unifdist(0, \, h(\theta^{(s)}))$ and accept $\theta^*$ if $V < h(\theta^*)$. 
This can be averaged with respect to the proposal distribution by first drawing $\theta^* \sim \widehat{\Pi}$, followed by $V$ to simulate the acceptance decision. 
This average is the probability that the state will change (as opposed to rejecting a proposal and remaining at $\theta^{(s)}$). 
The procedure and acceptance condition with a simple slice sampler are exactly the same but with the order reversed. 
First, draw $V \sim \Unifdist(0, \, h(\theta^{(s)}))$ to define the slice region. 
Given $V=v$, draw $\theta^* \sim \widehat{\Pi}$ and accept if $\theta^*$ falls in the slice region $\{ \theta^* : v < h(\theta^*) \}$.

Viewed from the IMH perspective, the joint density of $V$ and $\theta^*$ is 
\begin{align*}
p(\theta^*, v) = \frac{\hat{\pi}(\theta^*)}{h(\theta^{(s)})}1 \{ \theta^* \in \mathcal{S}_{\hat{\pi}}, \, 0 \le v \le h(\theta^{(s)}) \} \, .    
\end{align*}
The probability of a move is calculated by integrating this joint density over the region $R_{\text{IMH}} = \{ 0 < v < \min \left( h(\theta^*), \, h(\theta^{(s)}) \right) \}$, provided the integral exists. 
With the simple slice sampler, the expected probability of the slice region is
\begin{align}
    \label{eq:sliceprob}
    \E_{V}\left[ \widehat{\Pi} \left( \{ \theta^* : v < h(\theta^*) \} \mid \theta^{(s)}, V = v \right) \right] &= \int_{0}^{h(\theta^{(s)})} \widehat{\Pi} ( \{ \theta^* : v < h(\theta^*) \} \mid \theta^{(s)}, V = v ) \frac{1}{h(\theta^{(s)})} \diff v \nonumber \\
    &= \iint_{R_{\text{SS}}} \frac{\hat{\pi}(\theta^*)}{h(\theta^{(s)})} \diff \theta^* \diff v \, ,
\end{align}
where $R_{\text{SS}} = \{ 0 < v < h(\theta^{(s)}) \} \cap \{ v < h(\theta^*) \} = R_{\text{IMH}}$. 
The integrands and regions of integration are identical for IMH and the slice sampler. 
Both integrals are equal to 
\begin{align*}
\E_{\widehat{\Pi}}[ \alpha_{h}(\theta^{(s)}, \theta^*) \mid \theta^{(s)} ] = \frac{\hat{\pi}(\theta^{(s)})}{\pi(\theta^{(s)})} \widehat{\Pi}\left( \{ h(\theta^*) < h(\theta^{(s)}) \} \right) + \widehat{\Pi}\left( \{ h(\theta^*) \ge h(\theta^{(s)}) \} \right) \, ,
\end{align*}
thus, quantities Q1 and Q2 are equal.

To establish equality between quantities Q2 and Q3, 
begin with the expected probability of the slice region in \eqref{eq:sliceprob}, repeated here for convenience: 
\begin{align}
    \label{eq:sliceprob2}
    \E_{V}\left[ \widehat{\Pi} \left( \{ \theta^* : v < h(\theta^*) \} \mid \theta^{(s)}, V = v \right) \right] &= 
        \int_{0}^{h(\theta^{(s)})} \int_{\mathbb{R}}  \frac{\hat{\pi}(\theta^*)}{h(\theta^{(s)})} 1\{0 < v < h(\theta^*)\} \diff \theta^* \diff v \, .
\end{align}
Consider the change of variables $Y^* = \widehat{\Pi}(\theta^*) \in [0,1]$ and $U = V h_\psi(\psi^{(s)}) / h(\theta^{(s)}) \in (0, \, h_\psi(\psi^{(s)}))$. 
The Jacobian of the transformation has determinant $[\hat{\pi}(\widehat{\Pi}^{-1}(Y^*)) h_\psi(\psi^{(s)})]^{-1} h(\theta^{(s)})$. 
Let 
\begin{align*}
 a \defeq \frac{h(\theta^*) h_{\psi}(\psi^{(s)})}{h(\theta^{(s)})} = \frac{h_\psi(y^*) h_{\psi}(\psi^{(s)})}{h_\psi(\psi^{(s)})} = h_\psi(y^*) \, .
\end{align*}
The integral in \eqref{eq:sliceprob2} under transformation becomes
\begin{align}
    \label{eq:sliceprobTrans}
    & \quad \: \int_{0}^{h_\psi(\psi^{(s)})} \int_{0}^1  \frac{1}{h_\psi(\psi^{(s)})} 1\{0 < u < a \} \diff y^* \diff u \nonumber \\
    & =\int_{0}^{h_\psi(\psi^{(s)})} \left[ \int_{0}^1  1\{0 < u <  h_\psi(y^*) \} \diff y^* \right] \frac{1}{h_\psi(\psi^{(s)})} \diff u \nonumber \\
    &=\int_{0}^{h_\psi(\psi^{(s)})} w(u) \frac{1}{h_\psi(\psi^{(s)})} \diff u \, ,
\end{align}
where $w(u)$ is the total length of the slice region. 
The final integral \eqref{eq:sliceprobTrans} is precisely the inner integral in the first line of \eqref{eq:avg_sliceWidth} giving the expected total slice width. 
Thus, quantities Q2 and Q3 are equal. 
Therefore, all three quantities are equal. 

To show equality of $\E_{\Pi} \big[ \E_{\widehat{\Pi}}( \alpha_{h}(\theta^{(s)}, \theta^*) ) \, \big]$ and $\E_h[\E(W \mid \psi^{(s)})]$, we first simplify notation by rewriting $\E_{\widehat{\Pi}}( \alpha_h(\theta^{(s)}, \theta^*) )$ as $\E(\alpha_h \mid \theta^{(s)})$, which we have shown is equal to $\E(W \mid \psi^{(s)})$. 
We again change variables with $\psi^{(s)} = \widehat{\Pi}(\theta^{(s)})$ to obtain   
\begin{align*}
    \E_\Pi \left[ \E(\alpha_h \mid \theta^{(s)}) \right] &= \int_{\mathcal{S}_\pi} \E(\alpha_h \mid \theta^{(s)}) \, \pi(\theta^{(s)}) \diff \theta^{(s)} \\
    &= \int_{\mathcal{S}_{\hat\pi}} \E(\alpha_h \mid \theta^{(s)}) \, \frac{ \pi\left( \widehat{\Pi}^{-1}(\psi^{(s)}) \right) }{ \hat{\pi}\left( \widehat{\Pi}^{-1}(\psi^{(s)}) \right) } \diff \psi^{(s)} \\
    &= \int_{\mathcal{S}_{\hat\pi}} \E(\alpha_h \mid \theta^{(s)}) \, c_g \, h_\psi( \psi^{(s)} ) \, \diff \psi^{(s)} \\
    &= \int_{\mathcal{S}_{\hat\pi}} \E(W \mid \psi^{(s)}) \, c_g \, h_\psi( \psi^{(s)} ) \, \diff \psi^{(s)} \\
    &= \E_h [ \E(W \mid \psi^{(s)}) ] \, ,
\end{align*}
where the density of the stationary distribution of $\psi^{(s)}$ is $c_g \, h_\psi(\psi^{(s)})$. 
\hfill $\square$

%% file: appx_simulation_details.tex
\section{Simulation studies: additional details}
\label{sec:appendix_sim}

The \texttt{coda} package \citep{coda_package} estimates the effective sample size for a time series of MCMC samples $\{x^{(s)}: s= 1,\ldots,S \}$ as $S \,\widehat{\text{Var}}(\{x^{(s)}\}) / \hat{v}_0(\{x^{(s)}\})$. 
Here, $\widehat{\text{Var}}(\cdot)$ is the standard sample variance, and $\hat{v}_0(\cdot)$ is an estimate of the power spectral density of an autoregressive (AR) model at frequency 0, given by $\hat{\sigma}_\epsilon^2 / (1 - \sum_{r=1}^R \hat\phi_r)^2$,
where $\hat{\sigma}_\epsilon^2$ and $\{\hat\phi_r\}$ are the estimated error variance and coefficients of an AR model of order $R$.

All benchmarking reported in Sections \ref{sec:compare_methods} and \ref{sec:DHR_comparison} was performed on AMD EPYC 7502 servers running 
Ubuntu 22.04.4 LTS (GNU/Linux 5.15.0 x86\_64) at 2.50 GHz with 128 CPUs. 
Code was run using R version 4.4.1 \citep{Rteam}. 
Each run (MCMC chain) was restricted to a single thread, with 10 jobs running in parallel. 

Benchmarking reported in Section \ref{sec:gprior} was performed on an Intel\,\textsuperscript{\tiny\textregistered} Xeon\,\textsuperscript{\tiny\textregistered} Platinum 8592+ server running 
Ubuntu 24.04.2 LTS (GNU/Linux 6.8.0-57 x86\_64) at 1.9 GHz with 256 CPUs. 
Code was run using R version 4.4.3. 
Each run was restricted to a single thread, with 20 jobs running in parallel. 

In all cases, server were monitored to ensure that no computationally intensive processes occupied the free CPUs. 
Independent replicate runs were scheduled in a randomized order.

\subsection*{Standard targets in Section \ref{sec:compare_methods}}

With five standard targets (see log-transformed gamma and inverse-gamma below), 13 samplers, and 100 replicate runs, the complete experiment consisted of 6,500 total runs that were completed in less than 30 minutes. 

At 5\% significance, no more than nine (of 100) Kolmogorov-Smirnov (K-S) tests rejected thinned chains from any sampler (null hypothesis of being distributed according to the respective target). Across all samplers and targets, the median and mean K-S rejection rates were 5\% and 4.9\% respectively. 

Table \ref{tab:standard_targets_tuning} reports the settings for each of the 13 tuned samplers compared in Section \ref{sec:compare_methods}. All chains were initialized at 0.2, within regions of high density for all targets, and run for 50,000 iterations. Procedures for finding approximations to the inverse-gamma target excluded 20 degrees of freedom due to avoid problems with tail discrepancy.

\begin{table}[tb]
    \centering
    \begin{tabular}{l rrr}
    \toprule
     & \multicolumn{3}{c}{Target} \\
     \cmidrule{2-4}
     Sampler & Normal & Gamma & Inverse Gamma \\
     \midrule
    Random walk & $c = 2.5$ & $c = 4$ & $c = 7$ \\
    Stepping out \& shrinkage & $w = 2.5$ & $w = 6$ & $w = 1.5$ \\
    Generalized elliptical & $t(0, \, 1, \, 20)$ & $t(2, \, 1.5, \, 1)$ & $t(0.5, \, 0.4, \, 1)$ \\
    Latent slice & $r = 0.05$ & $r = 0.05$ & $r = 0.02$ \\
    Qslice: MSW & $t(0, \, 0.98, \, 20)$ & $t(1.74, \, 1.69, \, 5)1\{x > 0\}$ & $t(0.41, \, 0.38, \, 1)1\{x > 0\}$ \\
    Qslice/IMH: AUC & $t(0, \, 1, \, 20)$ & $t(1.47, \, 1.82, \, 5)1\{x > 0\}$ & $t(0.34, \, 0.41, \, 1)1\{x > 0\}$ \\
    Qslice: MSW--samples & varies & varies & varies \\
    Qslice: AUC--samples & varies & varies & varies \\
    Qslice/IMH: AUC--diffuse & $t(0, \, 4, \, 20)$ & $t(1.47, \, 7.27, \, 5)1\{x > 0\}$ & $t(0.34, \, 1.66, \, 1)1\{x > 0\}$ \\
    Qslice: Laplace--Cauchy & $t(0, \, 1.58, \, 1)$ & $t(1.5, \, 2.21, \, 1)1\{x > 0\}$ & $t(0.33, \, 0.17, \, 1)1\{x > 0\}$ \\
    Qslice: MM--Cauchy & varies & varies & varies \\
     \bottomrule
    \end{tabular}
    \caption{Settings for tuned samplers in the timing simulation. In the random walk sampler, $c$ is the standard deviation of the normal proposal distribution. In the step-and-shrink slice sampler, $w$ is the stepping size. In all methods using pseudo-targets, $t(m, s, d)1\{x \in A\}$ represents a Student-$t$ distribution with location $m$, scale $s$, degrees of freedom $d$, and truncation to set $A$. In the latent slice sampler, $r$ is a rate parameter.}
    \label{tab:standard_targets_tuning}
\end{table}

\begin{figure}[p]
    \centering
    \includegraphics[width=6.5in]{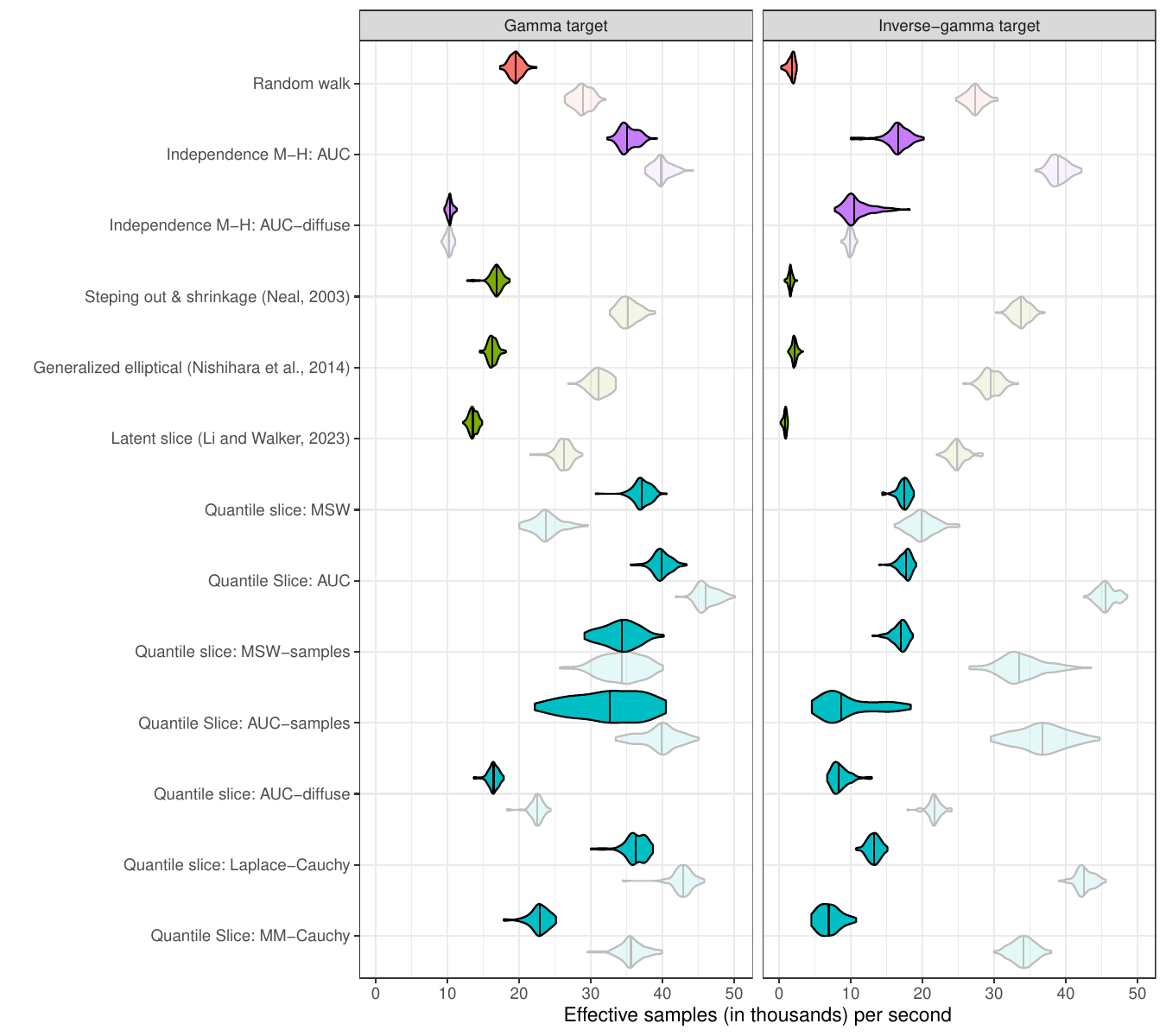}
    \caption{Violin plots summarizing effective samples per CPU second for 100 replicate MCMC chains of various samplers on standard targets with nonnegative support. Samplers are grouped by algorithm family and pseudo-target specification method. Light-colored densities show performance of each sampler on the target corresponding with a log-transformed random variable. Vertical lines indicate median effective samples per second.}
    \label{fig:sim_standard_targets_gammas}
\end{figure}

Figure \ref{fig:sim_standard_targets_gammas} repeats the simulation results for the gamma and inverse-gamma targets, but also includes results with each sampler applied (and tuned) to the target of a log-transformed gamma/inverse-gamma random variable. This common strategy for sampling on nonnegative support helps symmetrize the target to improve performance of standard samplers. Indeed, this is the case for the RWM and standard slice samplers, which outperform the less-optimized the Qslice/IMH samplers on the original gamma target. Logarithmic transformation improves \emph{all} samplers with the inverse-gamma target, as the resulting target is better approximated with a Student-$t$ distribution.

\subsection*{DHR simulation in Section \ref{sec:DHR_comparison}}
The MSlice, latent, and particle samplers were each tuned using pilot runs. 
Timing runs at selected settings consisted of two chains, each with 2,000 burn-in iterations followed by 20,000 iterations (both stages were increased five-fold for the MSlice and latent runs). 

%% file: appx_gprior.tex
\section{Hyper g-prior: additional details}
\label{sec:appendix_gprior}

With the model specification in Section \ref{sec:gprior}, the full conditional distributions for $\gamma$, $\sigma^{-2}$, and $\bm{\beta}$ are given as
\begin{align}
    \label{eq:fc_gprior}
    \pi(\gamma \mid \bm{\beta}, \sigma^{2}, \bm{y}) &\propto \gamma^{-p/2} (1 + \gamma)^{-a/2} \exp\left( \frac{-1}{2 \sigma^2 \gamma} \bm{\beta}^\top X^\top X \bm{\beta} \right) 1\{0 < \gamma < 3p^2\} \, , \\
    \sigma^{-2} \mid \bm{\beta}, \gamma, \bm{y} &\sim \Gammadist \left(a_\sigma + \frac{n + p}{2}, \, \text{rate}= b_\sigma + \frac{1}{2}\bm{\epsilon}^\top \bm{\epsilon} + \frac{1}{2\gamma} \bm{\beta}^\top X^\top X \bm{\beta} \right) \, , \nonumber \\
    \bm{\beta} \mid \gamma, \sigma^{2}, \bm{y} &\sim \Ndist \left( \frac{\gamma}{1+\gamma} \hat{\bm{\beta}} , \, \frac{\gamma \sigma^2}{1+\gamma} (X^\top X)^{-1} \right) \, , \nonumber
\end{align}
where $\bm{\epsilon} = \bm{y} - X \bm{\beta}$ and $\hat{\bm{\beta}} = (X^\top X)^{-1} X^\top  \bm{y}$ and $\Gammadist(a,b)$ denotes a gamma distribution with mean $a/b$. 

Five adaptive tuning rounds were used for each replicate run of the random walk (tuning $c$), and the step and shrink ($w$) and latent slice ($r$) samplers. In each round, 1,000 iterations were run using each of five different values of the tuning parameter. The tuning value yielding highest ESpS was retained for the subsequent round, along with four other values spanning half the range used in the previous round. The tuning-parameter value yielding the highest ESpS at the end of the fifth round was used for the 50,000 timing iterations.

Samplers using the AUC - samples tuning method set the pseudo-target from the final 2,000 samples of $\gamma$ from the burn-in period and proceeded directly to the timing iterations. Samplers using the analytical Laplace approximation proceeded immediately from burn-in to timing iterations with no tuning.

The analytical Laplace approximation to the full conditional for $\gamma$ is obtained using the first two derivatives of the log-full conditional:
\begin{align}
    \ell'(\gamma) = \frac{\diff}{\diff \gamma} \log(\pi(\gamma \mid \cdots)) &=  \frac{\bm{\beta}^\top X^\top X \bm{\beta}}{2\sigma^2\gamma^2} - \frac{a}{2(1+\gamma)} - \frac{p}{2\gamma} \, , \nonumber \\
    \ell''(\gamma) =\frac{\diff^2}{\diff \gamma^2} \log(\pi(\gamma \mid \cdots)) &= - \frac{\bm{\beta}^\top X^\top X \bm{\beta}}{\sigma^2 \gamma^3} + \frac{a}{2(1+\gamma)^2} + \frac{p}{2\gamma^2} \, . \nonumber
\end{align}
The location for the approximating Student-$t$ pseudo-target is set equal to the maximizing value $\hat{\gamma}$ obtained from the larger of two solutions to the quadratic equation $\ell'(\hat{\gamma}) = 0$. The scale of the pseudo-target is set equal to $(-\ell''(\hat{\gamma}))^{-1/2}$. 

An analogous procedure yields the Laplace approximation for the full conditional of $\Upsilon \defeq \log(\gamma)$ with 
\begin{align}
    \frac{\diff}{\diff \Upsilon} \log(\pi_{\Upsilon}(\Upsilon \mid \cdots)) &=  \frac{\bm{\beta}^\top X^\top X \bm{\beta}}{2\sigma^2 e^{\Upsilon}} - \frac{a e^{\Upsilon}}{2(1 + e^{\Upsilon)}} - \frac{p-2}{2} \, , \nonumber
\end{align}
the roots of which are obtained by solving a quadratic equation in $\gamma$. 
The Laplace--wide scale inflation factor is 20\% when sampling $\log(\gamma)$. 

%% file: appx_proofs_multivariate.tex
\section{Proof of Proposition \ref{prop:multivariate_unif_ergodicity}}
\label{sec:multivariate_unif_ergod}

We first extend the definition of shrinking intervals to shrinking hyperrectangles. 
Lemma \ref{lemma:reverse_hypercube_shrink} then establishes reversibility of the multivariate shrinkage procedure given in Figure 8 of \citet{neal2003slice} when initialized on the unit hypercube and applied to targets with support within the unit hypercube. 
The argument is very similar to the reversibility proof for Proposition \ref{prop:generalShrink}, but is restricted to shrinkage with a uniform full conditional as in \citet{neal2003slice}, who does not explicitly prove reversibility of the multivariate shrinkage procedure. 
Finally, we prove Proposition \ref{prop:multivariate_unif_ergodicity} with an argument that is nearly identical to the proof of Proposition \ref{prop:shrink_in_slice}.

\begin{definition}
    \label{def:shrink_hyperrect}
    Let $\{ \bm{z}_j : j = 0, 1, \ldots\}$ denote a countable sequence of real-valued vectors $\bm{z}_j = (z_{j,1}, \ldots, z_{j,D}) \in (0,1)^D$ that satisfies the following. 
    Let $\bm{z}_j \in \mathcal{H}_j(\bm{z}_{0:(j-1)}) \defeq (L_{j,1}, R_{j,1}) \times (L_{j,2}, R_{j,2}) \times \cdots \times (L_{j,D}, R_{j,D})$ for $j \ge 1$ with $\bm{z}_{0:(j-1)} \defeq (\bm{z}_0, \ldots, \bm{z}_{j-1})$, and $\{ (L_{j,d}, \, R_{j,d}) \}$ are recursively defined as follows. 
    Let $(L_{1,1}, \ldots, L_{1,D}) = (0, \ldots, 0)$ and $(R_{1,1}, \ldots, R_{1,D}) = (1, \ldots, 1)$. 
    Then, $(L_{j,d}, \, R_{j,d}) \defeq (z_{j-1,d}, \, R_{j-1,d})$ if $z_{j-1,d} < z_{0,d}$ and $(L_{j-1,d}, \, z_{j-1,d})$ otherwise, for all $d = 1, \ldots, D$, and $j > 1$.
\end{definition}

Let $K_{\Unifdist(0,1)^D, A}$ denote the transition kernel for the shrinkage procedure in Figure 8 of \citet{neal2003slice} targeting acceptance set $A$, but with the initial hyperrectangle fixed at the unit hypercube. 
Also let $\Unifdist(0,1)^D_A$ denote the uniform distribution on the $D$-dimensional unit hypercube restricted to set $A \subseteq (0,1)^D$ with positive Lebesgue measure $\lvert A \rvert$. 
To simplify notation, we use $\int_A f(\bm{z}) \diff \bm{z}$ to represent a $D$-dimensional integral over $A$.

\begin{lemma}
    \label{lemma:reverse_hypercube_shrink}
    The Markov transition kernel $K_{\Unifdist(0,1)^D, A}$ described above is $\Unifdist(0,1)^D_A$-reversible.
\end{lemma}

\noindent \textbf{Proof}:
We establish reversibility of $K_{\Unifdist(0,1)^D, A}$ by showing that it satisfies detailed balance. 
Let $B_0 \in \mathcal{B}((0,1)^D)$ and $B_1 \in \mathcal{B}((0,1)^D)$. 
Detailed balance requires that \\ $\int_{B_0} \lvert A \rvert^{-1} K_{\Unifdist(0,1)^D,A}(\bm{x}, B_1) \diff \bm{x} = \int_{B_1} \lvert A \rvert^{-1} K_{\Unifdist(0,1)^D,A}(\bm{x}, B_0) \diff \bm{x}$. 
The transition kernel can be represented as $K_{\Unifdist(0,1)^D,A}(\bm{x}, B) = \sum_{i=1}^\infty \Pr(\bm{x}_1 \in A \cap B \ \text{on $i$th step} \mid \bm{x})$ where, using the sequence from Definition \ref{def:shrink_hyperrect} and setting $\bm{z_0} = \bm{x}$,
\begin{align}
    \label{eq:shrink_jointprob_mv}
    \Pr(\bm{x}_1 &\in A \cap B \ \text{on $i$th step} \mid \bm{x}) = \Pr(\bm{z}_1 \notin A, \bm{z}_2 \notin A, \ldots, \bm{z}_{i-1} \notin A, \bm{z}_i \in A \cap B \mid \bm{z}_0) \nonumber \\
    &= \int_{\mathbb{R}^D} \ldots \int_{\mathbb{R}^D} \left[ \prod_{j=1}^{i-1} \frac{I\{ \bm{z}_j \in \mathcal{H}_j(\bm{z}_{0:(j-1)}) \setminus A \}}{ \lvert \mathcal{H}_j(\bm{z}_{0:(j-1)}) \rvert} \right] \frac{I\{ \bm{z}_i \in \mathcal{H}_i(\bm{z}_{0:(i-1)}) \cap A \cap B \}}{ \lvert \mathcal{H}_i(\bm{z}_{0:(i-1)}) \rvert} \diff \bm{z}_1 \ldots \diff \bm{z}_i \, .
\end{align}
Each of these terms in the summand operates on the same sequence $\{ \bm{z}_j \}$, with corresponding $\{ \mathcal{H}_j(\bm{z}_{0:(j-1)}) \}$. 
Only one term will be active for any given sequence. 

The key insight for establishing reversibility is that for any sequence $(\bm{x}, \bm{z}_1, \ldots, \bm{z}_{i-1}, \bm{z}_i)$ yielding positive transition density, the corresponding sequence of shrinking hyperrectangles will be \textit{identical} to the sequence of hyperrectangles arising from $(\bm{z}_i, \bm{z}_1, \ldots, \bm{z}_{i-1}, \bm{x})$. 
This is because all rejected points lie within the set $\{ (u_1, \ldots, u_D) : u_d < \min(x_d, z_{i,d}) \cup u_d > \max(x_d, z_{i,d}), \, d = 1, \ldots, D \}$, which is symmetric in $\bm{x}$ and $\bm{z}_i$, causing identical shrinking steps. 
Furthermore, for any $(\bm{x}, \bm{z}_i)$ pair, both $\bm{x}$ and $\bm{z}_i$ are contained in every hyperrectangle in $\{\mathcal{H}_j(\bm{z}_{j-1}) : j = 1, \ldots,i \}$. 
Letting 
\begin{align*}
    r_i(\bm{z}_{0:(i-1)}) \defeq \prod_{j=1}^{i-1} \frac{I\{ \bm{z}_j \in \mathcal{H}_j(\bm{z}_{0:(j-1)}) \setminus A \}}{\lvert \mathcal{H}_j(\bm{z}_{0:(j-1)})\rvert} 
\end{align*}
and $\bm{z}_0 = \bm{x}$, we have
\begin{align}
     \int_{B_0} & \lvert A \rvert ^{-1} K_{\Unifdist(0,1)^D, A}(\bm{x}, B_1) \diff \bm{x}  \nonumber \\
     &= \int_{B_0} \frac{I\{\bm{x} \in A \}}{\lvert A \rvert} \sum_{i=1}^\infty \int_{B_1} \int_{\mathbb{R}^D} \cdots \int_{\mathbb{R}^D} r_i(\bm{z}_{0:(i-1)}) \frac{I\{ \bm{z}_i \in \mathcal{H}_{i}(\bm{z}_{0:(i-1)}) \cap A  \}}{\lvert \mathcal{H}_i(\bm{z}_{0:(i-1)}) \rvert} \diff \bm{z}_1 \cdots \diff \bm{z}_i \diff \bm{x} \, \nonumber \\
     &=  \int_{B_0} \sum_{i=1}^\infty  \int_{B_1} \int_{\mathbb{R}^D} \cdots \int_{\mathbb{R}^D} r_i(\bm{z}_{0:(i-1)}) \frac{ I\{ \bm{x}, \bm{z}_i \in \mathcal{H}_{i}(\bm{z}_{0:(i-1)}) \cap A \}}{\lvert \mathcal{H}_i(\bm{z}_{0:(i-1)}) \rvert}  \frac{1}{\lvert A \rvert} \diff \bm{z}_1 \cdots \diff \bm{z}_i \diff \bm{x} \, \nonumber \\
     &= \sum_{i=1}^\infty  \int_{B_0} \int_{B_1} \int_{\mathbb{R}^D} \cdots \int_{\mathbb{R}^D} r_i(\bm{z}_{0:(i-1)}) \frac{ I\{ \bm{x}, \bm{z}_i \in \mathcal{H}_{i}(\bm{z}_{0:(i-1)}) \cap A \}}{ \lvert \mathcal{H}_i(\bm{z}_{0:(i-1)}) \rvert}  \frac{1}{\lvert A \rvert} \diff \bm{z}_1 \cdots \diff \bm{z}_i \diff \bm{x} \, \nonumber \\
     &= \sum_{i=1}^\infty  \int_{B_1} \int_{B_0} \int_{\mathbb{R}^D} \cdots \int_{\mathbb{R}^D} r_i(\bm{z}_{0:(i-1)}) \frac{ I\{ \bm{x}, \bm{z}_i \in \mathcal{H}_{i}(\bm{z}_{0:(i-1)}) \cap A \}}{ \lvert \mathcal{H}_i(\bm{z}_{0:(i-1)}) \rvert}  \frac{1}{\lvert A \rvert} \diff \bm{z}_1 \cdots \diff \bm{z}_i \diff \bm{x} \, \nonumber \\
     &=  \int_{B_1} \sum_{i=1}^\infty  \int_{B_0} \int_{\mathbb{R}^D} \cdots \int_{\mathbb{R}^D} r_i(\bm{z}_{0:(i-1)}) \frac{ I\{ \bm{x}, \bm{z}_i \in \mathcal{H}_{i}(\bm{z}_{0:(i-1)}) \cap A \}}{\lvert \mathcal{H}_i(\bm{z}_{0:(i-1)})\rvert}  \frac{1}{\lvert A \rvert} \diff \bm{z}_1 \cdots \diff \bm{z}_i \diff \bm{x} \, \nonumber \\
     &= \int_{B_1} \frac{I\{\bm{x} \in A \}}{\lvert A \rvert} \sum_{i=1}^\infty \int_{B_0} \int_{\mathbb{R}^D} \cdots \int_{\mathbb{R}^D} r_i(\bm{z}_{0:(i-1)}) \frac{I\{ \bm{z}_i \in \mathcal{H}_{i}(\bm{z}_{0:(i-1)}) \cap A  \}}{\lvert \mathcal{H}_i(\bm{z}_{0:(i-1)})\rvert} \diff \bm{z}_1 \cdots \diff \bm{z}_i \diff \bm{x} \, \nonumber \\
     &= \int_{B_1} \lvert A \rvert ^{-1} K_{\Unifdist(0,1)^D, A}(\bm{x}, B_0) \diff \bm{x} \, . \nonumber
\end{align}
\noindent We can interchange the order of summation and integration by Tonelli's theorem. 
Swapping the regions of integration in the central steps is justified because all shrinking intervals are identical and both $x$ and $z_i$ belong to the final interval.
Thus, $K_{\Unifdist(0,1)^D, A}$ is $\Unifdist(0,1)^D_A$-reversible.

\hfill $\square$

\noindent \textbf{Proposition \ref{prop:multivariate_unif_ergodicity}}: Assume target density $\pi(\bm{x}) \propto \tilde{h}(\bm{x}) I\{ \bm{x} \in (0,1)^D \}$ with lower-semicontinuous function $\tilde{h}: (0,1)^D \mapsto [0, \infty)$. 
Consider a simple slice sampler that defines the slice region with $V \mid \bm{x} \sim \Unifdist(0, \tilde{h}(\bm{x}))$ and uses the shrinkage procedure in Figure 8 of \citet{neal2003slice}, initialized on the unit hypercube, as a hybrid step targeting $p(\bm{x} \mid V=v) \propto I\{ \bm{x} \in A_{\tilde{h}}(v) \}$ with $A_{\tilde{h}}(v) \defeq \{ \bm{x} : v < \tilde{h}(\bm{x}) \}$. 
The transition kernel for this hybrid slice sampler is $\pi$-reversible. 
If $\sup_{\bm{x}} \tilde{h}(\bm{x}) < \infty$, the resulting Markov chain is uniformly ergodic.

\noindent \textbf{Proof}: 
\citet{latuszynski2024hybrid} showed (their Lemma 1) that reversibility of an embedded hybrid step with respect to the uniform distribution over $A_{\tilde{h}}(v)$ for all $v \in (0, \, \sup_{\bm{x}}{\tilde{h}(\bm{x})})$ is sufficient for the reversibility of a uniform slice sampler using that hybrid step. 
Reversibility of the uniform shrinkage procedure established in Lemma \ref{lemma:reverse_hypercube_shrink} for any measurable $A$ implies reversibility of uniform slice samplers using the multivariate shrinkage procedure described in the statement of the proposition. 

Ergodicity depends on the target. 
We next show that if $\tilde{h}$ is bounded, then uniform slice sampling on target $\tilde{h}(\bm{x})$ with the shrinkage procedure yields a uniformly ergodic Markov chain. 
First, consider $A = \cup_i A_i$, a possibly countable union of mutually disjoint hyperrectangles contained in the unit hypercube. 
Given the current point $\bm{x}_0$, the probability that the final accepted point $\bm{x}_1$ belongs to hyperrectangle $A_i$ is 
\begin{align}
    \label{eq:shrinkage_prob_subinterval_mv}
    \Pr(\bm{x}_1 \in A_i \mid \bm{x}_0, A) 
    &\ge \Pr(\text{first proposed} \ \bm{x}_1 \in A_i \mid \bm{x}_0, A) \\
    &= \lvert A_i \rvert \nonumber
\end{align}
where $\lvert A_i \rvert$ is the Lebesgue measure (generalized ``volume") of $A_i$. 
The first line takes only the first of countably many mutually exclusive outcomes tracking the proposal on which $\bm{x}_1$ accepted. 
The final equality recognizes that the first proposal is uniform over $(0,1)^D$. 

We use the probabilities from \eqref{eq:shrinkage_prob_subinterval_mv} to characterize the density of the finally accepted $\bm{x}_1$, given $A$ and $\bm{x}_0$. 
Conditional on $\bm{x}_1 \in A_i$, the density of $\bm{x}_1$ is uniform over $A_i$. 
Therefore,
\begin{align}
    \label{eq:uniform_shrink_density_mv}
    \begin{split}
        p_{\bm{x}_1}(\bm{x}_1 \mid \bm{x}_0, A) &= \sum_i \frac{1}{\lvert A_i \rvert} \Pr(\bm{x}_1 \in A_i \mid \bm{x}_0, A) I\{ \bm{x}_1 \in A_i \} \\
        &\ge \sum_i \frac{\lvert A_i \rvert}{\lvert A_i \rvert} I\{ \bm{x}_1 \in A_i \} \\
        &= \sum_i I\{ \bm{x}_1 \in A_i \} \, ,
    \end{split}
\end{align}
where the inequality uses \eqref{eq:shrinkage_prob_subinterval_mv}. 
Thus, the density of the accepted point is bounded below by 1 everywhere in $A$. 

We can now establish uniform ergodicity of the uniform slice sampler with shrinkage and support on the unit hypercube (USSUH) by verifying Doeblin's condition. 
Specifically, we show that the sampler is aperiodic and that the entire support of $\pi(\bm{x})$ is a small set. 
Following the strategy for Theorem 6 in \citet{mira2002sliceEfficiency}, we write the transition kernel of the slice sampler, $K_\text{USSUH}$, integrating over the slice-defining latent variable $V$. 
Note that the sets $A_{\tilde{h}}(v) = \{ \bm{x} : v < \tilde{h}(\bm{x}) \}$ can be expressed as a countable union of disjoint hyperrectangles, as assumed in \eqref{eq:shrinkage_prob_subinterval_mv} and \eqref{eq:uniform_shrink_density_mv}. 
With $\pi(\bm{x}) \propto \tilde{h}(\bm{x}) I\{ \bm{x} \in (0,1)^D \}$ and $\sup_{\bm{x}} \tilde{h}(\bm{x}) < \infty$, the single-step transition density from $\bm{x}_0 \in (0,1)^D$ to $\bm{x}_1 \in (0,1)^D$ of $K_\text{USSUH}$ is 
\begin{align}
    \label{eq:unif_ergodicity_unif_slice_mtv}
    \begin{split}
        k_\text{USSU}(\bm{x}_0, \bm{x}_1) &= \int_0^{\tilde{h}(\bm{x}_0)} \frac{1}{\tilde{h}(\bm{x}_0)} p_{\bm{x}_1}(\bm{x}_1 \mid \bm{x}_0, A_{\tilde{h}}(v)) \diff v \\
        &= \frac{1}{\tilde{h}(\bm{x}_0)} \int_0^{\min(\tilde{h}(\bm{x}_0), \, \tilde{h}(\bm{x}_1))} p_{\bm{x}_1}(\bm{x}_1 \mid \bm{x}_0, A_{\tilde{h}}(v)) \diff v \\
        &\ge \frac{1}{\tilde{h}(\bm{x}_0)} \int_0^{\min(\tilde{h}(\bm{x}_0), \, \tilde{h}(\bm{x}_1))} 1 \diff v \\
        &= \frac{\min(\tilde{h}(\bm{x}_0), \, \tilde{h}(\bm{x}_1))}{\tilde{h}(\bm{x}_0)} \\
        &\ge \frac{\tilde{h}(\bm{x}_1)}{\sup_{\bm{x}}{\tilde{h}(\bm{x})}} \\
        &= \frac{c_{\tilde{h}}}{\sup_{\bm{x}}{\tilde{h}(\bm{x})}} \pi(\bm{x}_1) \, ,
    \end{split}
\end{align}
where $c_{\tilde{h}} = \int_{(0,1)^D} \tilde{h}(\bm{x}) \diff \bm{x}$. 
Thus $\pi(\bm{x})$ itself, which is positive on all of its support by assumption, supplies the minorizing measure.
\hfill $\square$

%% file: appx_DHR.tex
\section{Dynamic harmonic regression: implementation}
\label{sec:appendix_DHR}

The time-varying parameters have the following initial distributions, 
    $\beta_{0,0} \sim \mathcal{N}(0, 5^2)$, 
    $\alpha_{1,0} \sim \mathcal{N}(0, 4^2) \, 1\{\alpha_{1,0} \ge 0\}$, and 
    $\alpha_{2,0} \sim \mathcal{N}(0, 2^2) \, 1\{\alpha_{2,0} \ge 0\}$, 
and evolve in the prior according to random walks with continuous-time adjustment 
\begin{align*}
  \beta_{0,t} \mid \beta_{0,t-1} &\sim \mathcal{N}(\beta_{0,t-1}, \, 0.03[s_t - s_{t-1}]) \, , \\
  \alpha_{1,t} \mid \alpha_{1,t-1} &\sim \mathcal{N}(\alpha_{1,t-1}, \, 0.1875[s_t - s_{t-1}]) \, 1\{\alpha_{1,t} \ge 0\} \, , \ \text{and} \\
  \alpha_{2,t} \mid \alpha_{2,t-1} &\sim \mathcal{N}(\alpha_{2,t-1}, \, 0.1875[s_t - s_{t-1}]) \, 1\{\alpha_{2,0} \ge 0\} \, ,
\end{align*}
with $1\{\cdot\}$ indicating truncation to enforce positivity constraints. 
The model is completed with uniform priors on $\phi_1$ and $\phi_2$, and an inverse-gamma prior on $\sigma^2$ with shape $5/2$ and scale $5 \times 0.25/2$. 

We use a Gibbs sampler that cycles among updates for the observation variance $\sigma^2$, the phase vector $(\phi_1, \phi_2)$, time-varying intercepts $\{ \beta_{0,t} : t = 1, \ldots, T\}$ (as a block), and amplitude vectors $\balpha_1 = (\alpha_{1,1}, \ldots, \alpha_{1,T})$ and $\balpha_2$. Note that each time-varying parameter (TVP) block (vector) is updated separately to facilitate working with truncated distributions. The only exception is the full FFBS implementation that ignores parameter constraints and updates all TVPs together.

In Gaussian DLMs, the FFBS algorithm exploits the Markovian prior and conditional independence to decompose the joint full conditional for $\bm{\alpha}_j$ into cascading univariate densities
\begin{align}
\label{eq:ffbs}
p(\bm{\alpha}_j \mid y_{1:T}, \cdots ) = p(\alpha_{j,T} \mid y_{1:T}, \cdots) \prod_{t = T-1}^1 p(\alpha_{j,t} \mid \alpha_{j,t+1}, y_{1:t}, \cdots) \, ,
\end{align}
where $y_{1:t}$ is understood to mean all observations up to time index $t$. 
The conditional distributions in \eqref{eq:ffbs} are Gaussian provided the initial distributions, state evolution distributions, and observation distributions for $y_t$ are linear and Gaussian. 

We construct a joint pseudo-target for $\bm{\alpha}_j$ from cascading univariate conditional densities, as in Section \ref{sec:pseu-condseq}. 
The pseudo-target distributions are truncated Gaussian or Student-$t$ with location and scale parameters obtained from the corresponding distributions in \eqref{eq:ffbs}. 
The transformation $\bm{\psi}_j = (\widehat{\Pi}_T(\alpha_{j,T}), \, \widehat{\Pi}_{T-1}(\alpha_{j, T-1} \mid \alpha_{j,T}), \, \ldots, \, \widehat{\Pi}_1(\alpha_{j,1} \mid \alpha_{j,2}, \ldots, \alpha_{j,T}) )$ yields a one-to-one mapping from $(0, \infty)^T$ to $(0, 1)^T$ on which we apply the multivariate slice sampler with shrinkage. 

The update for $\sigma^2$ is a simple Gibbs draw from the conjugate full conditional. 
The phases are updated using a bivariate quantile slice sampler with independent pseudo-targets. Each pseudo-target is a modular, symmetric Beta$(2,2)$ distribution with mode at the previous $\phi$ value and wrapping, that is, $\lim_{\phi \rightarrow 0^+} \hat\pi(\phi) = \lim_{\phi \rightarrow 1^-} \hat\pi(\phi)$.

Because the intercepts in $\{ \beta_{0,t} : t = 1, \ldots, T\}$ are unconstrained, we use with a standard FFBS step applied to the mean-adjusted data
\begin{align*}
    \tilde{y}_{0,t} &\defeq y_t - \alpha_{1,t} \cos\left(2\pi[s_t + \phi_1]\right) - \alpha_{2,t} \cos\left(2\pi[2s_t + \phi_2]\right) \\
    &\simindep \Ndist(\beta_{0,t}, \, \sigma^2)\, .    
\end{align*}
Full conditional distributions for amplitude vectors $\balpha_j$ ($j = 1,2$) are obtained using similarly adjusted observations. In the case of the annual cycle ($j=1$), we have 
\begin{align}
\label{eq:fc_alpha}
\begin{split}
    \tilde{y}_{1,t} &\defeq y_t - \beta_{0,t} - \alpha_{2,t} \cos\left(2\pi[2s_t + \phi_2]\right) \\ 
    & \simindep \Ndist \left( \alpha_{1,t} \cos\left(2\pi[s_t + \phi_1]\right), \, \sigma^2 \right) \, .
\end{split}
\end{align}
The observation densities in \eqref{eq:fc_alpha} are combined with densities for the initial and evolution distributions for $\{ \alpha_{1,t} \}$ in Section \ref{sec:DHR_model} to define the target full conditional used by the competing methods, as well as the FFBS-based pseudo-target used by the quantile slice sampler. The update using the quantile slice sampler is given in Algorithm \ref{alg:ampltiude_quantile_slice}.

The particle MCMC sampler implements Algorithms 16.5, 16.7, and 16.8 of \citet{chopin2020book}. It uses a set number of particles plus a so-called ``star trajectory" that represents the previous value of $\balpha_j$. The MSlice and latent slice algorithms use common width (rate) tuning parameters that influence the size of the initial shrinking hyperrectangle. Test runs of MSlice with time-specific widths (influenced by the FFBS variances, for example) did not result in clear improvement of algorithm performance.

\begin{algorithm}[p]
    \caption{Quantile slice update for an amplitude vector.} 
    \label{alg:ampltiude_quantile_slice}
    \begin{algorithmic}
        \State \textbf{Input:}
        \State $\balpha^{\text{in}}$: Previous state of $\balpha = (\alpha_1, \ldots, \alpha_T) \in (0,+\infty)^T$
        \State $f$: Target density function; product of observation and state densities
        \State $\tilde{\bm{y}}$: Vector of (mean-adjusted) observations
        \State $\Theta$: Parameters governing the observation and state distributions
        \State $\mathcal{P}$: Pseudo-target family and auxiliary parameters (degrees of freedom, for example)
        \State \textbf{Output:}
        \State $\balpha^{\text{out}}$: New state for $\balpha$
        
\medskip
\hrule
\medskip

        \State Run unconstrained forward filter with $\Theta$ and $\tilde{\bm{y}}$ to obtain location and scale parameters $\{ \mu_t^{\text{FF}} : t = 1, \ldots, T\}$ and $\{ \sigma_t^{\text{FF}} \}$. \\

        \State Use $\balpha^{\text{in}}$, $\{ \mu_t^{\text{FF}} \}$, $\{ \sigma_t^{\text{FF}} \}$, and $\Theta$ to obtain unconstrained backward-sampling location and scale vectors $\{ \mu_t^{\text{BS}} \}$ and $\{ \sigma_t^{\text{BS}} \}$. \\

        \State Define the pseudo-target as the sequence of conditional distributions, truncated at 0 and ordered from $t=T$ to $t=1$, with locations in $\{ \mu_t^{\text{BS}} \}$ and scales in $\{ \sigma_t^{\text{BS}} \}$. The joint pseudo-target has density $\hat{\pi}(\balpha)$ with vector-valued (sequential) CDF $\widehat{\Pi}(\balpha) \in (0,1)^T$ and vector-valued inverse-CDF $\widehat{\Pi}^{-1}(\bm{\psi}) \in (0,+\infty)^T$ for $\bm{\psi} \in (0,1)^T$. \\
        
        \State Draw $v \sim \Unifdist(0,f(\balpha^{\text{in}})/\hat{\pi}(\balpha^{\text{in}}))$ \Comment{Define the slice region.} \\

        \State $\bm{L} \defeq (L_1, \ldots, L_T) \leftarrow (0, \ldots, 0)$
        \State $\bm{R} \defeq (R_1, \ldots, R_T) \leftarrow (1, \ldots, 1)$ \\

        \State $\bm{\psi}^{\text{in}} \defeq (\psi_1^{\text{in}}, \ldots, \psi_T^{\text{in}}) \leftarrow \widehat{\Pi}(\balpha^{\text{in}})$ \Comment{Transform to probability space.} \\

        \Loop
        \For{$t$ in $1:T$}
            \State $\psi_t^\star \sim \Unifdist(L_t, R_t)$ \Comment{New proposal drawn uniformly between $L_t$ and $R_t$.}
        \EndFor
        
        \State $\balpha^\star \leftarrow \widehat{\Pi}^{-1}(\bm{\psi}^\star)$ \Comment{Proposal on original state space.}
        
        \If{$f(\balpha^\star)/\hat{\pi}(\balpha^\star) > v$} \textbf{break}
        \Else
            \For{$t$ in $1:T$}
                \If{$ \psi_t^\star < \psi_t^{\text{in}}$} $L_t \leftarrow \psi_t^\star$
                \Else~$R_t \leftarrow \psi_t^\star$
                \EndIf                
            \EndFor

        \EndIf
        \EndLoop
        
        \State $\balpha^{\text{out}} \leftarrow \balpha^\star$
    \end{algorithmic}
\end{algorithm}